\documentclass[aps,prd,groupedaddress,floatfix,longbibliography,superscriptaddress]{revtex4-2}

\usepackage{amssymb}
\usepackage{xcolor}
\usepackage{subfig}
\usepackage{float}
\usepackage{braket}
\usepackage{slashed}

\usepackage{graphicx}
\usepackage{amsmath}
\usepackage{soul}
\usepackage[normalem]{ulem}

\newcommand{\gv}[1]{\ensuremath{\mbox{\boldmath$ #1 $}}}

\newcommand{\beq}{\begin{equation}}
\newcommand{\eeq}{\end{equation}}

\newcommand{\barr}{\begin{eqnarray}}
\newcommand{\earr}{\end{eqnarray}}

\newcommand{\mbf}{\mathbf}

\begin{document}

%\today

\title{Lorentz invariance violation and the CPT-odd electromagnetic response of a tilted anisotropic Weyl semimetal} 

%\today

\author{Andrés G\'omez}
\email{andresgz@ciencias.unam.mx}
\affiliation{Facultad de Ciencias, Universidad Nacional Aut\'{o}noma de M\'{e}xico, 04510 Ciudad de M\'{e}xico, M\'{e}xico}
\affiliation{Institut f\"{u}r Theoretische Physik, Universit\"{a}t Heidelberg, Philosophenweg 16, 69120 Heidelberg, Germany}

\author{R. Mart\'\i nez von Dossow}
\email{ricardo.martinez@correo.nucleares.unam.mx}
\address{Instituto de Ciencias Nucleares, Universidad Nacional Aut\'{o}noma de M\'{e}xico, 04510 Ciudad de M\'{e}xico, M\'{e}xico}

\author{A. Mart\'{i}n-Ruiz}
\email{alberto.martin@nucleares.unam.mx}
\address{Instituto de Ciencias Nucleares, Universidad Nacional Aut\'{o}noma de M\'{e}xico, 04510 Ciudad de M\'{e}xico, M\'{e}xico}

\author{Luis F. Urrutia}
\email{urrutia@nucleares.unam.mx}
\affiliation{Instituto de Ciencias Nucleares, Universidad Nacional Aut\'{o}noma de M\'{e}xico, 04510 Ciudad de M\'{e}xico, M\'{e}xico}

\begin{abstract}

We derive the electromagnetic response of a particular fermionic sector in the minimal QED contribution  to the Standard Model Extension (SME), which can be physically realized in terms of a  model describing   a tilted and anisotropic Weyl semimetal (WSM). The contact is  made through the  identification of  the Dirac-like Hamiltonian resulting from the SME  with that corresponding to the WSM in the linearized tight-binding approximation. We first  calculate the effective  action by computing the nonperturbative vacuum polarization tensor using thermal field theory techniques, focusing upon the corrections at finite chemical potential and zero temperature.
Next, we confirm our results  by a direct calculation of the anomalous Hall current within a chiral kinetic theory approach.
 In an ideal Dirac cone picture of the WSM (isotropic and non-tilted) such response is known to be governed by axion electrodynamics, with the space-time dependent axion angle $\Theta ({\gv r},t) = 2 ({\gv b} \cdot {\gv r} - b _{0} t)$, being $2 {\gv b}$ and $2b _{0}$ the separation of the Weyl nodes in momentum and energy, respectively. In this paper we demonstrate that the node tilting and the anisotropies induce novel corrections at a finite density which however preserve the structure of the axionic field theory. 
We apply our results to the ideal Weyl semimetal  $\mbox{EuCd}_{2}\mbox{As}_{2}$ and to the highly anisotropic and tilted monopnictide TaAs.

\end{abstract}

\maketitle

\section{Introduction}

The one loop  effective action of QED in terms of external electromagnetic fields  is a powerful tool to study multiphoton  interactions at energies  where the fundamental fermions  are not excited, as well as transport properties arising from the resulting  currents  \cite{Dittrich:2000zu,Dittrich:1985yb}. This action is obtained  by ``integrating'' the fermions,  yielding  an effective Lagrangian which introduces additional contributions to the Maxwell term having the general form of  a non-linear  electromagnetic response. A distinguished  member of this class is the Euler-Heisenberg Lagrangian \cite{Heisenberg:1936nmg} 
{which correctly anticipated  some important results in QED, among which we find: (i) light-light scattering, as first discussed in Ref. \cite{Euler:1935zz} and subsequently given a full solution in Ref. \cite{Karplus:1950zz}; (ii) pair-production from vacuum in an electric field, already noticed in \cite{Heisenberg:1936nmg} motivated in part from 
Ref. \cite{Sauter:1931zz} and  later given a complete description in \cite{Schwinger:1951nm} and   (iii) the need of  charge renormalization,  further developed in Refs. \cite{Schwinger:1954zza, Schwinger:1954zz}}.
This effective Lagrangian is known to all orders in the electromagnetic field only in a restricted family of backgrounds, such as  constant fields or plane wave fields, to name some well known cases. For a review see for example 
\cite{Dunne:2004nc}.  The extension to the case of non-homogeneous electromagnetic fields has remained a  subject of investigation and several advances have been reported \cite{Berestetskii:1982qgu,Narozhnyi:1970uv,Martin1988,Dunne:1997kw,Dunne:1998ni}. 
The inclusion of temperature and density, which is relevant to the study of transport properties in many body theory (e.g. metals, topological matter and  quark-gluon plasma) and in  spontaneously broken theories, has also been undertaken \cite{Chodos:1990vv,Weldon:1982aq,Ahmed:1991mz,le_bellac_1996,Kapusta:2006pm,Dittrich:1979ux,Bernard:1974bq,Das:1997gg,kubo_statistical_1999}. A recent review can be found in  Ref. \cite{Mustafa:2022got}.
 
  The  advent of high power lasers together with increasingly energetic particle beams has fostered the theoretical and experimental interest in the general theory of QED with intense background fields. Higher intensity, together with higher accuracy, demands an increase in the number of loops in the calculation and also to develop further nonperturbative methods. A detailed review of this whole topic, with emphasis on the advances in  the last decade, is presented in Ref. \cite{Fedotov:2022ely}. Among the new theoretical tools providing an alternative to the standard Feynman diagrams calculations, the semiclassical worldline instanton method, deriving from  the ``first quantized'' approach to field theory, has proven particularly useful in the calculation of effective actions and related quantities in the study  of QED processes in external fields \cite{Schubert:2001he,Schubert:2023bed}.

Effective QED actions  have been extensively studied in the framework of the Standard Model Extension (SME) \cite{PhysRevD.55.6760,PhysRevD.58.116002}. 
 This model parametrizes  Lorentz invariance  violations (LIV) in the fundamental interactions and by itself can be viewed as an effective model  resulting from a more fundamental theory where this symmetry could be spontaneously broken by nonzero vacuum expectation values (VEVs) of tensorial operators. Contrary to the scalar Higgs field, such VEVs  introduce fixed directions in space-time  yielding the presumed violation. Such fixed tensors are coupled to the fields of the Standard Model  providing all possible violating terms in the Lagrangian, consistent with the known symmetries of the fundamental interactions. These parameters must be extremely suppressed in order to  find agreement with current experimental observations.

 %%%%%%%%%%%%%%%%%%%%%%%%%%%%%%%%
A relevant issue in this framework is to find relations  among the actions characterizing  a given sector of the model, in order to reduce the proliferation of coefficients which codify the LIV, as well as to understand the induced radiative corrections. This provides a natural setup for searching new effective electromagnetic actions arising from the additional couplings of the fermions to the electromagnetic field. 
A much studied particular case includes  the CPT-odd violating terms in the fermion-photon sector of the minimal QED extension of the SME. There we find  the modified  Dirac action  
\begin{align}
 S_{\rm D} = \int d ^{4} x \, {\bar \Psi}(x) \left( i \gamma ^{\mu} \partial _{\mu} - {{\tilde b}_{\mu}  \gamma ^{5}  \gamma ^{\mu} } \right) \Psi(x), \label{1}
\end{align}
which includes the LIV coefficient   {${\tilde b}^{\mu}$}, together with the {CPT-odd}  contribution to the photon sector
\begin{align}
S _{{\rm{eff}}} = \frac{e ^{2}}{32 \pi ^{2}} \int d ^{4}x \,  \Theta(x) \,  \epsilon^{\alpha \beta \mu \nu} F _{\alpha \beta} F_{\mu \nu}, \quad \Theta(x) = {\tilde c}_\mu x ^{\mu},  \label{02},
\end{align}
codified by another  LIV coefficient ${\tilde c}_\mu$. Incidentally, the addition of $(\ref{02})$ to the standard Maxwell action yields Carroll-Field-Jackiw (CFJ) electrodynamics \cite{PhysRevD.41.1231}, 
 which can also be considered as a restricted version of axion-electrodynamics
 \cite{Peccei:1977hh,Sikivie:1983ip} since the axion field $\Theta$ is not dynamical. The challenge here is to regain the action (\ref{02}) through radiative corrections induced by the fermionic coupling in (\ref{1}), which  effectively amount to obtain the corresponding  effective electromagnetic  action. The conclusion is that in fact  ${\tilde c_\mu} =\zeta {\tilde b_\mu} $, but with $\zeta$ being undetermined,  having  a finite value which depends on the regularization method. This was the subject of intense debate in the literature as can be appreciated by the numerous references to the topic.
 A survey of the principal approaches  regarding this issue can be found in Refs.  \cite{Jackiw:1999yp, PhysRevLett.83.2518, Battistel:2000ms, Perez-Victoria:2001csb,Andrianov:2001zj,Altschul:2003ce,Altschul:2004gs, Alfaro:2009mr,Jackiw:1999qq,Altschul:2019eip}. 
 Additional work yielding the effective one loop electromagnetic action induced by many of the additional LIV terms appearing in the fermionic sector of the minimal QED extension of the SME include: 
nonperturbative \cite{Furtado:2014cja, Mariz:2007gf, Ferrari:2021eam} and perturbative calculations for some specific LIV parameters to first order \cite{Mariz:2016ooa, Anacleto:2014aha, BaetaScarpelli:2013rmt, Casana:2013nfx, Gomes:2009ch}, the inclusion of higher  derivative terms in the action \cite{Mariz:2010fm, Mariz:2011ed, Borges:2016uwl} and  higher order contributions of the LIV parameters in the one loop effective action
\cite{BaetaScarpelli:2018vsm}.  A review including  these results is found in  \cite{Ferrari:2018tps} and   references therein. 
%%%%%%%%%%%%%%%%%%%%%%%%%%%%%%%%%%%%%%%%%%%%%%%%

Recently, a very interesting connection between the SME and the area of topological materials in condensed matter, where LIV occurs naturally,  has been found   through  the identification of fermionic quasiparticles (excitations) of Dirac and/or Weyl type in the linearized approximation of tight-binding Hamiltonians of topological phases of matter  in regions close to the Fermi energy. Then, one can embed such Hamiltonians into a Dirac-Weyl field theory of the form (\ref{1}) and subsequently  apply the wealth of tools already developed in previous studies of LIV.
In particular, the electromagnetic transport properties can be obtained through the calculation of the effective electromagnetic action, computed with standard methods in field theory.  It is important to observe that, contrary to the case in high energy physics, the LIV parameters that one identifies in such materials need not be highly suppressed, but are determined by their electronic structure and further subjected to experimental determination. In this way, the standard perturbative approach  frequently used in most related calculations in high energy physics, might prove inadequate in the condensed matter case, where  nonperturbative methods could be required to obtain  realistic results.
Steps in this direction can be found in Refs. \cite{Grushin:2012mt,Kostelecky:2022idz, McGinnis:2023ubw, Urrutia:2022dxj, Gmez2022,Kostelecky:2021bsb}. 
 
 As we will show in the following, the  use of quantum field theory  methods can  be particularly fruitful in the case of Weyl semimetals (WSMs), whose electronic Hamiltonians naturally include some of the LIV terms considered in the fermionic sector of the SME.

{WSMs are topologically nontrivial conductors in which the valence and conduction bands touch at isolated points, (the so-called Weyl nodes) in the Brillouin zone, locally forming Dirac cones \cite{PhysRevLett.111.027201, Armitage:2017cjs,Yan:2016euz}. According to the Nielsen-Ninomiya theorem \cite{Nielsen:1983rb},  the Weyl nodes in crystals occur in pairs of opposite chirality, indicating the presence of fermionic excitations of the Weyl-type. The individual nodes within a pair act as a source and sink of the Berry curvature, a topological property of the electronic Bloch wave functions. In this way  the WSM phase is topologically protected by a nonzero Berry flux across the Fermi surface.}

{WSM phases in crystals require either broken spatial inversion symmetry or broken time-reversal symmetry, or both. The Weyl phase with broken inversion symmetry has been predicted \cite{10.1038/ncomms8373, PhysRevX.5.011029} and experimentally confirmed for the family of transition metal monopnictides compounds TaAs, NbAs, TaP, and NbP \cite{doi:10.1126/science.aaa9297, doi:10.1126/science.aaa9273, PhysRevX.5.031013, PhysRevLett.115.217601, 10.1038/nphys3437, 10.1038/nphys3426, 10.1038/nmat4457}. Those with broken time-reversal symmetry have been proposed for pyrochlore iridates R$_2$Ir$_2$O$_7$ (R is a rare-earth element) such as Y$_2$Ir$_2$O$_7$ \cite{PhysRevB.83.205101, PhysRevB.85.045124, PhysRevB.86.235129}.}

{In general, Weyl nodes in solids do not behave exactly in the same manner as their  high-energy Lorentz invariant analogs because they are tilted and anisotropic. In a noncentrosymmetric WSM, for example, the Weyl nodes appear in pairs of opposite chirality, opposite tilting and rotated anisotropy. These deviations from the ideal Dirac cone picture influences several properties of WSMs like optical \cite{10.1038/nmat4457}, spin texture \cite{PhysRevLett.116.096801}, and expectedly, several anomalous transport phenomena.}

{There are a few  common approaches for studying electronic transport in WSMs, including the Kubo formula \cite{doi:10.1143/JPSJ.12.570}, the chiral kinetic theory \cite{PhysRevLett.109.162001} and the  loop  effective action of quantum electrodynamics.   In order to compare our results obtained within a quantum field theory approach  we will use the chiral kinetic theory in this paper. This is a topologically modified semiclassical Boltzmann formalism  (SBF) to describe the behavior of Weyl fermions at finite density. Within this approach, the many-electron system is described by a moving wave packet whose center satisfies semiclassical equations of motion augmented by an anomalous velocity term arising from the Berry curvature, which acts as a magnetic field in reciprocal space. }

The relevance of the topological Berry curvature in the calculations of the transport properties, together with the presence of the abelian Pontryagin density as a factor in the effective action (\ref{02}) arising from ({\ref{1}) has promoted the use of tools akin to  anomaly calculations in field theory to obtain the effective action. Let us recall that  the electromagnetic chiral anomaly is proportional to the abelian Pontryagin density whose integral is a topological invariant,  thus suggesting that the topological properties ensuing from the Berry curvature  in the SBF could be understood as a manifestation of the anomaly in the field theory approach. For example, using  path integral methods the effective action (\ref{02}) was obtained  by introducing the electromagnetic coupling in  Eq.~(\ref{1}) and subsequently eliminating the fermionic term proportional to ${\tilde b}_\mu$ through a chiral rotation. Nevertheless, instead of yielding a free fermionic action this  produces an  electromagnetic contribution  arising from the nonzero Jacobian of the chiral transformation which is proportional to the Pontryagin density \cite{Zyuzin:2012tv}.
Following this idea, the Fujikawa prescription to obtain the chiral anomalies \cite{fujikawa2004path,bertlmann2000anomalies} has also been used to calculate the effective electromagnetic action of different materials  in Refs. \cite{Goswami:2012db,Sekine:2020ixs,Chernodub:2021nff}.
 Nevertheless, as pointed out in Refs. \cite{Arias:2007xt,Salvio:2008ta,BaetaScarpelli:2015ykr,Gomez:2021aez}
the anomaly does not incorporate all the parameters which one would expect to determine the full dynamics of  the effective action. Also,  the method of eliminating the additional  fermionic contributions via a chiral rotation  cannot be  easily extended to deal with the more complicated configurations envisaged  in generalizations of the action (\ref{1}). The reasons indicated above  point to the need of  presenting  a full quantum field theory method   to obtain  the required effective electromagnetic actions corresponding to fermionic systems described by such extensions.   This is the purpose of this work. In particular, this procedure  should clarify how  the LIV corrections enter in the effective action, while the chiral anomaly remains insensitive to them, as mentioned above. As an application, this method  will  provide us with an alternative way to calculate the electromagnetic response of some particular cases in topological quantum matter.

In this work we consider the more general fermionic action 
\begin{align}
S = \int d ^{4} x \, \bar{\Psi} \left( \Gamma ^{\mu} i \partial _{\mu} - M
- e \Gamma ^{\mu} A _{\mu} \right) \Psi,  \label{ACTION1}
\end{align}
 coupled to the electromagnetic field $A_\mu$, and we  restrict ourselves to
\begin{equation}
\Gamma ^{\mu }={ \gamma^\mu}+c^{\mu }{}_{\nu }\gamma ^{\nu }+d^{\mu
}{}_{\nu }\gamma ^{5} \gamma ^{\nu },\qquad M=a_{\mu }\gamma ^{\mu }+b_{\mu
}\gamma ^{5} \gamma ^{\mu }.
\label{GAMMAMU}
\end{equation}
 This   corresponds to a particular choice of coefficients in the SME where  we set  $m=m_5=H_{\mu\nu}=e_\mu=f_\mu=g_{\mu\nu\lambda}=0 $, in the notation of Table P52 of Ref. \cite{RevModPhys.83.11}. In Eq. (\ref{GAMMAMU}), the matrices   $\gamma ^{\mu }$ are the standard ones and the isolated $\gamma^\mu$ contribution takes care of Lorentz covariant piece of the   Dirac field. Also we have $\bar{\psi}=\psi^\dagger \gamma^0$. Our main motivation  for this choice is that, close to the Fermi energy,  the  linear approximation  of the tight-binding Hamiltonian of a WSM with two cones, having arbitrary tilting and anisotropy, can be embedded in the action (\ref{ACTION1}). This will provide the opportunity to obtain the electromagnetic response from standard field theory methods via the effective action, as well as  to compare these results with those obtained using the SBF. {Also, we focus upon the CPT-odd contribution
 to the electromagnetic response since this sector describes novel and interesting phenomena such as the anomalous Hall effect and also provides contributions to the chiral magnetic effect}.
 Since the conduction process strongly depends  on the fermion filling of the valence and conduction bands we require  to  consider  a finite particle density, which we achieve by introducing the chemical potential at zero temperature as a first approximation. 
 
Aside from further detailing the calculations in  our previous results \cite{Gmez2022}, together with extending them to the anisotropic case, our aim in this work is  to make the first steps in establishing   the relation between the two alternative methods considered; the effective action calculation and the semiclassical Boltzmann formalism. This  might shed additional light in the role that the  chiral anomaly plays in the characterization of the electromagnetic transport properties of WSMs  in the framework of the effective action.}

The paper is organized as follows. In section \ref{II}  we define the general effective electromagnetic action and consider only the restriction to the  CPT-odd contribution of the vacuum polarization tensor $\Pi^{\mu\nu}$,  which we relate to the resulting axionic electrodynamics describing the response of the medium we  study. The  action to be integrated in the presence of an external electromagnetic field is selected from the fermionic sector of the SME with the choices indicated in Eq. (\ref{GAMMAMU}) and turns out to be chiral. The detailed decomposition 
of the full vertices and propagators into their chiral contributions, labeled by $\chi= \pm 1$, is carried in section \ref{CHIRALPROY} together with Appendix \ref{APA}. In this way, the vacuum polarization tensor is split into two contributions,  $\Pi^{\mu\nu}_\chi$, whose expressions are analogous and are calculated in Section \ref{IIB} with the help of the Appendix \ref{APB}. The condensed matter Hamiltonian describing a tilted anisotropic WSM, whose electromagnetic response is obtained as an application  of the previous results, is introduced in Section \ref{III}. We show how to embed this Hamiltonian in the fermionic action (\ref{GAMMAMU}) and write the relations among their parameters. To calculate the effective current we incorporate the finite density regime via the chemical potential $\mu$ at zero temperature, which is introduced in section \ref{IV} using the Matsubara prescription. The calculation is further split into a  $\mu$-independent contribution, calculated in Section \ref{IVA} plus the Appendix \ref{APint}, together with  a $\mu$-dependent piece,  carrying  all the information regarding the tilting and anisotropy, which is summarized in Section \ref{IVB} and heavily relies on the Appendixes \ref{APC}, \ref{APD} 
and \ref{APE}. Section \ref{Kinetic_Theory_App} is devoted to the calculation of the anomalous Hall current  using the kinetic theory approach, as a way of comparing the effective action results with a well established and powerful method in condensed matter physics. Some applications  of the matching results focusing on the anomalous Hall current are described in Section \ref{APPL} using  $\mbox{EuCd}_{2}\mbox{As}_{2}$ and $\mbox{TaAs}$, which are well known tilted and anisotropic WSMs. Finally we close in section \ref{VII} with the summary and results. 
Our metric convention  is $\eta_{\mu\nu}={\rm diag}(1,-1,-1, -1)$ and $\epsilon^{0123}=+1$.

\section{ The effective action}

\label{II}

Let us start from the action (\ref{ACTION1}) together with the selection  (\ref{GAMMAMU}) for $\Gamma^\mu$ and $M$.
We are interested in a low energy regime where the fermions are not excited, yielding an effective contribution to the electromagnetic interaction  allowing the determination of the induced current. To this end we calculate 
the effective action $ S_{\rm eff}(A)$ given by
\begin{align}
    \exp [iS_{\rm eff}(A)] &= \int D\bar{\Psi}D\Psi \exp \left[  i \int d^{4}x\bar{\Psi}  \left( \Gamma ^{\mu } i \partial _{\mu} - M - e \Gamma ^{\mu} A _{\mu} \right) \Psi \right] = \det \left( \Gamma ^{\mu } i \partial _{\mu} - M - e \Gamma ^{\mu} A _{\mu} \right) . \label{DET}
\end{align}
Following the standard procedure we introduce  the noninteracting Green function $S={i}/{\left( i \Gamma^{\mu }\partial_{\mu }-M\right) }$
and we write
\begin{align}
    \det  \left( i\Gamma ^{\mu} \partial _{\mu} - M -e \Gamma ^{\mu} A _{\mu} \right)  = \det \left( \Gamma ^{\mu }i\partial _{\mu }-M\right) \det \left[ 1-S \, \Gamma ^{\alpha }\left( -ieA_{\alpha }\right) \right]. \label{DET1}
\end{align}
Discarding  the  irrelevant normalization factor\ $\det \left( i \Gamma
^{\mu }\partial_{\mu }-M\right) $, using the identity  $\det M=\exp \mathrm{Tr}\ln M$ and the power expansion of the logarithm we solve $S_{\rm eff}$ from Eq.(\ref{DET}) obtaining
\begin{align}
    iS_{\rm eff}(A)=  \, \mbox{Tr} \sum_{n=1}^{\infty }-\frac{1}{n}\left[  S \, \left( -ie\Gamma ^{\alpha }A_{\alpha }\right) \right] ^{n}. \label{SEFF1}
\end{align}
The trace $\textrm{Tr}$ is taken in coordinate as well as in  matrix space, while $\textrm{tr}$ is reserved to the trace in  matrix space.
To  second order in $A_{\alpha }$ we obtain 
\begin{eqnarray}
iS_{\rm eff}^{(2)}(A) 
&=&\frac{e^{2}}{2}\int d^{4}x\;d^{4}x^{\prime }\;A_{\mu }(x^{\prime })  \, \mbox{tr} 
\left[  S(x-x^{\prime }) \Gamma ^{\mu }
S(x^{\prime }-x) \Gamma ^{\nu }\right] A_{\nu }(x),
\label{EFF_ACT}
\end{eqnarray}%
in coordinate space. Going to the Fourier space, with the conventions
\begin{equation}
A_{\nu }(x)=\int \frac{d^{4}k}{\left( 2\pi \right) ^{4}}e^{-ikx}A_{\nu
}(k),\qquad  S(x-x^{\prime })=\int \frac{d^{4}k}{\left( 2\pi \right) ^{4}}%
e^{-ik\left( x-x^{\prime }\right) }S(k),\qquad i\partial _{\mu }=k_{\mu
},
\label{CONVFOURIER}
\end{equation}%
yielding
\beq
S(k)=\frac{i}{  \Gamma ^{\mu }k_{\mu }-M },
\label{PROP}
\eeq
we recast Eq. (\ref{EFF_ACT}) as%
\begin{equation}
iS_{eff}^{(2)}(A)=+\frac{e^{2}}{2}\int \frac{d^{4}p}{\left( 2\pi \right) ^{4}%
}A_{\mu }(-p)\left[ \int \frac{d^{4}k}{\left( 2\pi \right) ^{4}}  \, \mbox{tr} \left[
S(k-p)\Gamma ^{\mu }S(k)\Gamma ^{\nu }\right] \right] A_{\nu }(p).
\label{SEFF2}
\end{equation}%
Let us introduce the vacuum polarization tensor $\Pi ^{\mu \nu }(p)$  
\begin{equation}
i\Pi ^{\mu \nu }(p)=e^{2}\left[ \int \frac{d^{4}k}{\left( 2\pi \right) ^{4}}%
 \, \mbox{tr} \left[ S(k-p)\Gamma ^{\mu } S(k)\Gamma ^{\nu }\right] \right],
\label{PIMUNU1}
\end{equation}%
which produces the final expression for the effective action 
\begin{equation}
S_{\rm eff}^{(2)}(A)=\frac{1}{2}\int \frac{d^{4}p}{\left( 2\pi \right) ^{4}}%
A_{\mu }(-p)\;\Pi ^{\mu \nu }(p)\;A_{\nu }(p),  \label{DEFL}
\end{equation}
Since $S _{\rm eff} ^{(2)}$ is real we must have $\Pi^{*}_{\mu\nu}(p)=\Pi_{\nu\mu}(p)$. 

In the following we consider only the CPT-odd contribution to the effective action (\ref{DEFL}) which, as we will show,  keeps the form of Eq. (\ref{02}) with a new vector ${\cal B}_\lambda$ to be determined, { which replaces the original  ${\tilde c}_\lambda$.
The resulting vacuum polarization contribution together with the new axion field is }
\begin{align}
\Pi ^{\mu \nu} (p) = -i \frac{e ^{2}}{2 \pi ^{2}} \, {\cal B} _{\lambda} p _{\kappa} \epsilon ^{\mu \nu \lambda \kappa} , \quad \Theta (x) = 2 {\cal B} _{\lambda}  x ^{\lambda} . \label{BLAMBDA}
\end{align}
In other words, the  LIV parameters of the model $c_
{\mu\nu}, d_{\mu\nu}, a_\mu, b_\mu $  will contribute only  through {the vector } ${\cal B}_\lambda $ yielding the  full  electromagnetic action (in Gaussian units) 
\beq
S[A_\mu]= \int d^4x \Big[ -\frac{1}{16 \pi} F_{\mu\nu} F^{\mu \nu}-\frac{1}{c}  J^\mu A_\mu+ \frac{\alpha}{16\pi^2} \Theta(x) F_{\mu\nu}{ \tilde F}^{\mu\nu}  \Big] ,
\label{GU}
\eeq
where the electromagnetic tensor is  $F_{\mu\nu}=\partial_\mu A_\nu-\partial_\nu A_\mu$,  with its dual  ${ \tilde F}^{\mu\nu} =\frac{1}{2} \epsilon^{\mu\nu \alpha\beta}\, F_{\alpha \beta }$, and $\alpha=e^2/(\hbar c)$ is the fine structure constant. 
The resulting equations of motion are 
\beq
\partial_\mu F^{\mu\nu}=\frac{4 \pi}{c} J^\mu+
 \frac{\alpha}{\pi} (\partial_\mu \Theta ) { \tilde F}^{\mu\nu}.
 \label{EQMOT}
\eeq
Following the conventions of Ref. \cite{jackson_classical_1999}, we have
\barr
&& \hspace{-1cm}\gv{\nabla}\cdot \mbf{E}= 4\pi \rho+\frac{\alpha}{\pi} (\gv{\nabla} \Theta)\cdot \mbf{B},
\qquad 
 \gv{\nabla} \times \mbf{B}-\frac{1}{c}\frac{\partial \mbf{E}}{\partial t}= \frac{4 \pi}{c} \mbf{J}- \frac{\alpha}{\pi} \frac{1}{c} 
\frac{\partial \Theta}{\partial t} \mbf{B}- \frac{\alpha}{\pi} (\gv{\nabla} \Theta) \times \mbf{E},
\label{EQCOMP}
\earr
{in terms of the electromagnetic fields. Equations (\ref{EQCOMP}) yield} the effective current densities
\barr
&& \rho_{\rm eff}=\frac{\alpha}{4\pi^2} (\gv{\nabla} \Theta)\cdot \mbf{B}, \qquad \mbf{J}_{\rm eff}=-\frac{c \alpha}{4 \pi^2} \Big(
\frac{\partial \Theta}{\partial x^0} \mbf{B} + (\gv{\nabla} \Theta) \times \mbf{E}\Big).
\label{EFFCURR}
\earr
{which provide a realization of the magnetoelectric effect.  From the homogeneous Mawxwell's equations $\partial_\mu {\tilde F}^{\mu\nu}=0$ one can verify the effective charge conservation $\partial_t \rho_{\rm eff}+ \gv{\nabla} \cdot {\mbf J}_{\rm eff}=0$ for an arbitrary coordinate dependent axion field $\Theta(x)$.}
{The equations (\ref{EFFCURR}) }can be written in terms of ${\cal B}_0=\partial_0 \Theta/2$ and $\gv{\cal B}=\{ {\cal B}^i \}$, with $={\cal B}_i=-{\cal B}^i= \partial_i \Theta/2$. 
{The current $\mbf{J}_{\rm{AHE}}=(c\alpha/2\pi^2) \gv{\cal B}\times \mbf{E}$ describes the anomalous Hall effect, while $\mbf{J}_{\rm CME}=-(\alpha/2 \pi^2){\cal B}_0 \mbf{B}$ contributes to the chiral magnetic effect.} 
In the following we set $\hbar=c=1$ so that $\alpha = e ^{2} = 1/137$.

\subsection{The chiral propagators} 

\label{CHIRALPROY}

Now we give some preliminary steps for the calculation of the vacuum polarization tensor. In the absence of the unit matrix, our generalized Dirac operator $\Gamma ^{\mu} k _{\mu} - M $, in Eq.(\ref{GAMMAMU}), is linear in $\gamma ^{\mu}$ and $\gamma ^{5} \gamma ^{\mu}$. The appearance of the matrix $\gamma ^{5}$ suggests the convenience of using  left and right chiral projectors in order to replace $\gamma^5$ by its eigenvalues $\pm 1$.
{This procedure was first performed  in Ref.\cite{Altschul:2003ce} and subsequently used in Refs.  \cite{Salvio:2008ta, 10.1140/epjc/s10052-008-0677-4}, among others}.
Therefore, it is convenient to  
define the  operators
\begin{align}
P _{\chi} = \frac{ 1 + \chi \gamma ^{5} }{2},  \qquad \gamma _{5} ^{2} = 1,  \qquad P_++P_-=1, \qquad P_\chi^2=P_\chi, \qquad  P_+\, P_-=P_-\, P_+=0, \label{11}
\end{align}
which project onto the right-handed (R) and the  left-handed (L) subspace, with $\chi=+1$ and $\chi=-1$, respectively. Note that $\gamma ^{\mu} P _{\chi} = P _{-\chi} \gamma ^{\mu}$. The projectors (\ref{11}) allows us to define the matrices $\Gamma _{\chi} ^{\mu}$ such that
\begin{align}
\Gamma ^{\mu} P _{\chi} = \left({{\delta^\mu}_\nu}+c ^{\mu} {}_{\nu} - \chi \, d ^{\mu} {}_{\nu} \right) \gamma ^{\nu} P _{\chi} \equiv \Gamma _{\chi} ^{\mu} P _{\chi} , \label{12}
\end{align}
which explicitly identifies 
\barr
&&\Gamma^\mu_{\chi} = (m_\chi)^{\mu} {}_{\nu} \,  \gamma ^{\nu}, \qquad  (m_\chi)^{\mu} {}_{\nu}={{\delta^\mu}_\nu}+ c ^{\mu} {}_{\nu} -  \chi d ^{\mu} {}_{\nu}.
\label{GAMMACHI}
\earr
The apparent mismatch that $(m_\chi)^\mu{}_\nu$  gets a $-\chi$  factor in front of  $d ^\mu{}_\nu$ is readily clarified recalling that in our conventions  we have  $\gamma^5 \gamma^\mu P_{\chi}=-\gamma^\mu \gamma^5 P_{\chi}  = - (\chi) \gamma^\mu  P_{\chi} $. In the following we indistinctly use  the notation $\chi= (+1,-1) $ or $\chi= (R, L)$.

To proceed forward with the calculation we now split the combination $T ^{\mu \nu} (k,p) = S(k-p) \Gamma ^{\mu} S(k) \Gamma ^{\nu}$ inside  the trace of  $\Pi ^{\mu \nu }(p)$ in Eq. (\ref{PIMUNU1}) into its left- and right-handed parts, i.e. 
\begin{align}
T _{\chi} ^{\mu \nu } (k,p) &= S(k-p) \Gamma ^{\mu} S(k) \Gamma ^{\nu} P _{\chi} , \label{18}
\end{align}
which implies that the vacuum polarization can be written as the sum $\Pi ^{\mu \nu} (p) = \Pi _{L} ^{\mu \nu} (p) + \Pi _{R} ^{\mu \nu} (p)$, with 
\begin{align}
i \Pi _{\chi} ^{\mu \nu} (p) &= e ^{2} \int \frac{d ^{4} k}{\left( 2 \pi \right) ^{4}} \, \mbox{tr} \left[ T _{\chi} ^{\mu \nu} (k,p) \right]  \label{17} 
\end{align}
being the  vacuum polarization of a massless fermion with chirality $\chi$. 

The next step is  to calculate $S(k) \Gamma^\nu P_\chi$, with $S(k)$ given in Eq. (\ref{PROP}). The $\Gamma^\nu $ in the numerator changes into a linear combination of standard matrices $\gamma^\alpha$ according to  Eq. (\ref{12}),  but we are still left with $S(k)\gamma^\alpha P_\chi$ where we require to determine the action of the projector in the  denominator of the propagator.  This is done in the Appendix \ref{APA} with the results 
\beq
\frac{i}{\Gamma ^{\mu} k _{\mu} - M } \gamma ^{\alpha} P _{\chi} = P _{\chi} S _{\chi} (k) \gamma ^{\alpha}, \quad  S _{\chi} (k) = \frac{i}{\big( k _{\mu} (m_\chi)^{\mu}{}_{\nu} -{(C_\chi)}_{ \nu}  \big) \gamma ^{\nu}}, 
\qquad  {(C_\chi)}_{ \nu}= a_\nu-\chi b_\nu. \\  
\label{141}
\eeq
Note that the propagators $S_{\chi}$, having the generic form $i/(Z_\nu \gamma^\nu)$, can be readily rationalized as $i (Z_\nu \gamma^\nu)/ Z^2 $.

\subsection{The vacuum polarization tensor}

\label{IIB}

We now concentrate in the calculation of each contribution $\Pi_{\chi} ^{\mu \nu} (p)$. Inserting Eqs. (\ref{PIMUNU1}) and (\ref{141}) into (\ref{18}) and (\ref{17}) and using  the cyclic property of the trace we have 
\begin{align}
 i \Pi _{\chi} ^{\mu \nu} (p) &= e ^{2} (m_\chi) ^{\mu} {}_{\beta} \, (m_\chi)^{\nu} {}_{\alpha} \int \frac{d^{4}k}{\left( 2\pi \right) ^{4}} \, \mbox{tr} \left[ S _{\chi} (k-p) \gamma ^{\beta}  S _{\chi} (k) \gamma ^{\alpha} P _{\chi} \right] . \label{19}
\end{align} 
{Having in mind the application of our results to the transport properties of WSMs,}
in the following we restrict ourselves to the CPT-odd (axial) contributions  $\Pi ^{\mu \nu} _{A,\chi}$  of the vacuum polarization, which are obtained by selecting the terms  $ \chi  \gamma ^{5} /2$ in the projector $P_\chi$ of Eq. (\ref{19}). Then we are left with
\begin{align}
i \Pi _{A, \, \chi } ^{\mu \nu} (p) &= \frac{{\chi}}{2} e^{2} (m_\chi)^{\mu}{}_{\beta} (m_\chi) ^{\nu}{}_{\alpha} \int \frac{d ^{4} k}{\left( 2 \pi \right) ^{4}} \mbox{tr} \left[ S _{{\chi}} (k-p) \gamma ^{\beta} S _{{\chi}} (k) \gamma ^{\alpha} \gamma ^{5} \right] ,   \label{21}
\end{align}

Clearly, the full axial contribution to the vacuum polarization is the sum of the $L$ and $R$ parts, i.e. $ \Pi _{A} ^{\mu \nu} = \Pi _{A, L} ^{\mu \nu} + \Pi _{A, R} ^{\mu \nu}$. The calculation indicated in   Eq. (\ref{21}) is presented in the Appendix \ref{APB} with the result
\begin{align}
\Pi _{A, \, \chi }^{\mu \nu }(p) = - 2 {\chi} e ^{2} (\det m_\chi)(m_\chi^{-1}) ^{\rho} {}_{\lambda} \epsilon ^{\mu \nu \lambda \kappa} p _{\kappa} \, I^\chi _{\rho} (C),  \label{23}
\end{align}
with
\begin{align}
I^\chi _{\rho} (C) = \int  \frac{d ^{4} k }{\left( 2 \pi \right) ^{4}} \, g^\chi _{\rho} (k _{0} , {\gv k}) \, , \quad  g^\chi _{\rho} (k _{0} , {\gv k}) = \frac{ \left( k'_\chi - C_\chi \right) _{\rho} }{ \left[ \left( k'_\chi - C_\chi \right) ^{2}\right] ^{2} }, \label{24}
\end{align}
where we have taken that ${(k'_\chi)}_\mu={k}_\alpha \,  (m_\chi)^{\alpha}{}_\mu$. 
Previous to regularization, the above expression is our final result for the vacuum polarization in Minkowski spacetime.  The result (\ref{23}) holds for arbitrary LIV parameters $c^\mu{}_\nu, \, d^\mu{}_\nu, \, a_\mu$  and $b_\mu$ as long as these produce invertible matrices $(m_\chi)^{\mu}{}_\nu$. Note that this result is  nonperturbative in these parameters. 
{From Eqs. (\ref{BLAMBDA}) and (\ref{23})  we can read the chiral contributions to ${\cal B}_\lambda={\cal B}^+_\lambda+{\cal B}^-_\lambda$ as
\beq
{\cal B}^\chi_\lambda=-i4 \pi^2 \chi (\det  m_\chi) (m^{-1}_\chi)^\rho{}_\lambda\, I_\rho^\chi(C) .
\label{CHIRALB}
\eeq}

{

%\color{blue}

\section{The model}

\label{III}

In order to further motivate the additional choices we make to obtain our final result for ${\cal B}_\mu$ which will determine the  electromagnetic response (\ref{EFFCURR}), we introduce a simple model of a WSM consisting of two Weyl nodes of opposite chiralities separated in momentum and energy, ignoring the nonuniversal corrections due to band bending far away from the nodes. The low-energy Hamiltonian for a Weyl node with chirality $\chi$ can be expressed as 
\begin{align}
H _{\chi} (\gv{k}) = {\gv v} _{\chi} \cdot ({\gv k}   +  \chi {\gv {\tilde b}}) \sigma _{0}   -   \chi {\tilde b}_{0} \sigma _{0} + \chi  ({\gv k} + \chi {\gv {\tilde b}}){\mathbb A} _{\chi}   {\gv \sigma}, \qquad {\gv W} {\mathbb A}_\chi \gv{\sigma}\equiv W ^{i} A _{\chi\, ij } \gv{\sigma}^j ,  \label{Hamiltonian_Tilting}
\end{align}
where ${\gv k}$ is the crystal momentum, $\boldsymbol{\sigma}$ is the vector of Pauli spin matrices, $\sigma_0$ is the $2 \times 2$ unit matrix, and $W^i$ denotes an arbitrary vector. This model describes two Weyl nodes located at $- \chi \tilde{\textbf{b}}$ ($-   \chi {\tilde b}_{0}$) {in momentum (energy) with respect to}  the origin at ${\gv k} = {\gv 0}$ (the zero-energy plane). 
The tilting velocity of each cone is ${\gv v} _{\chi} = \{ v _{\chi} ^{i} \} $ and ${\mathbb{A}} _{\chi} = [(A _{\chi}) _{ij}]$ is the matrix of anisotropic Fermi velocities, with the notation indicated in the second term of Eq. (\ref{Hamiltonian_Tilting}).

The dispersion relation of this model is
\begin{align}
E _{s \chi } ( {\gv k} ) = - \chi {\tilde b}_{0}  + {\boldsymbol{\mathcal{V}}} _{\chi} \cdot {\boldsymbol{\mathcal{K}}} _{\chi} + s \mathcal{K} _{\chi} , \label{Energy}
\end{align}
where $s = \pm 1 $ is the band index,  ${\boldsymbol{\mathcal{V}}} _{\chi} =  {\mathbb A} _{\chi} ^{-1} {\gv v} _{\chi} $, ${\boldsymbol{\mathcal{K}}} _{\chi} = ({\gv k} + \chi {\gv {\tilde b}}){\mathbb A} _{\chi}$ and $\mathcal{K} _{\chi} = \vert {\boldsymbol{\mathcal{K}}} _{\chi}  \vert $.  In order to illustrate the emergence of the linearly dispersing model we are considering {in Eq.} (\ref{Hamiltonian_Tilting}), the left panel of Fig. \ref{Fig_Dispersion} shows a general energy dispersion for a two-node Weyl semimetal, including band bending far away from the nodes {described by a more general tight-binding Hamiltonian}.  When the Fermi level is close to the band crossings, {this} Hamiltonian can be linearized around each node, yielding to our model of Eq. (\ref{Hamiltonian_Tilting}).  In the middle panel of Fig. \ref{Fig_Dispersion} we show the Weyl cones without tilting and {with} isotropic Fermi velocity.  The right panel displays tilted Weyl cones with anisotropic Fermi velocity.

\begin{figure}
    \centering
    \includegraphics[width=0.28\textwidth]{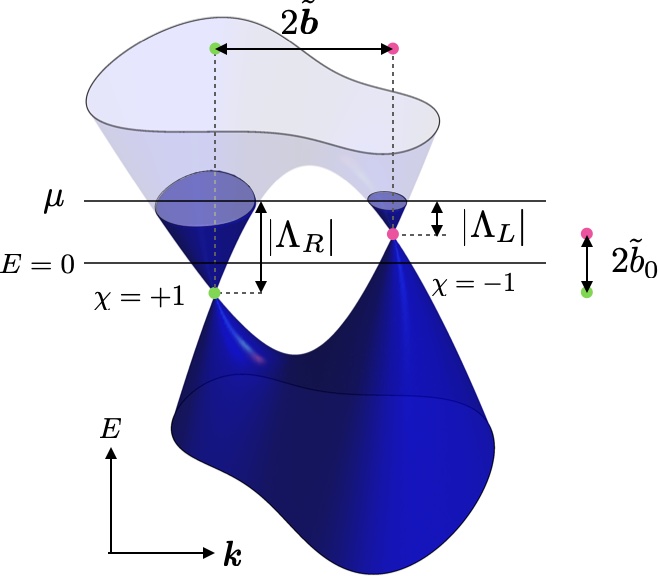} \hspace{0.2cm}
    \includegraphics[width=0.32\textwidth]{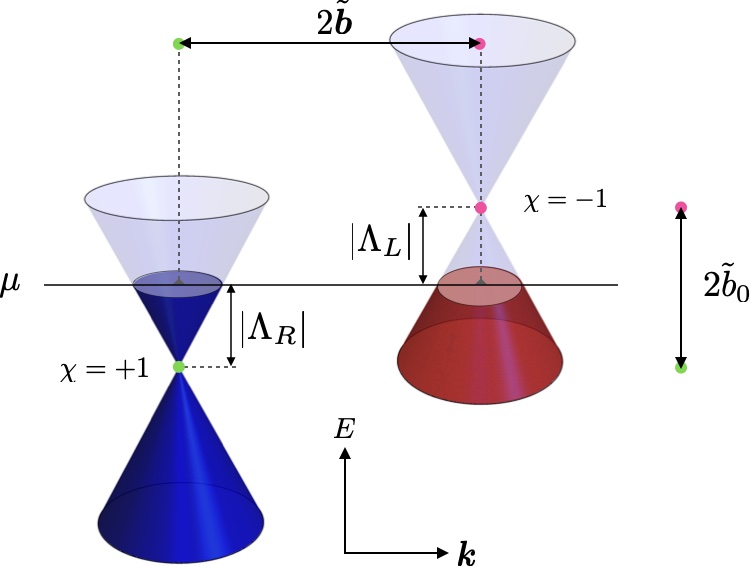} \hspace{0.2cm} \includegraphics[width=0.32\textwidth]{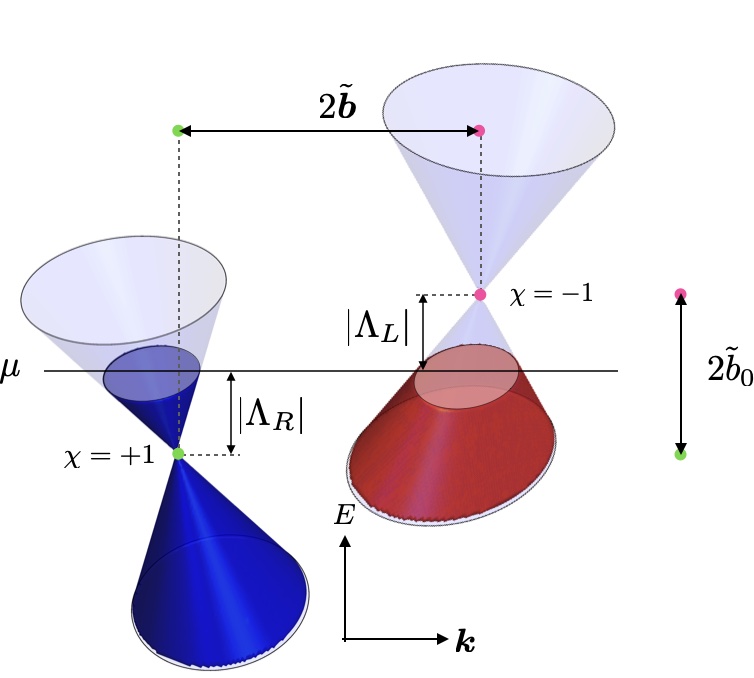}
    \caption{Left: energy dispersion for a two-node WSM  including band bending far away from the nodes.   Middle: Low-energy spectrum of a two-node WSM without tilting and {with } isotropic Fermi velocity.  Right: Low-energy spectrum of a two-node tilted Weyl cones with anisotropic Fermi velocity. The black line represents the position of the Fermi level $\mu$ and $\Lambda _{\chi} = \mu + \chi \tilde{b} _{0}$ measures the band filling. } \label{Fig_Dispersion}
\end{figure}

An important step in our calculation is the identification of the {parameters in the condensed matter Hamiltonian  (\ref{Hamiltonian_Tilting}) with  the parameters  entering  the free Hamiltonian, i.e. without the electromagnetic coupling, resulting from  the fermionic action (\ref{ACTION1})  in the SME. These relations allow us to express the results obtained from the effective action in Section \ref{II}, in terms of the physical parameters characterizing the WSM.}
To facilitate this identification  we rewrite (\ref{Hamiltonian_Tilting}) as
\begin{align}
    H _{L} ({\gv k}) &= {\gv v}_L \cdot {\gv k}  -  {\gv v}_L \cdot {\gv {\tilde b}}  +  {\tilde b}_{0} - {\gv k}{\mathbb A}_L  {\gv \sigma} + {\gv {\tilde b}}{\mathbb A}_L  {\gv \sigma},  \label{CML}  \\ H _{R} ({\gv k}) &=  {\gv v}_R \cdot{\gv k}   + {\gv v}_R \cdot{\gv {\tilde b}}  -  {\tilde b}_{0} +  {\gv k}{\mathbb A}_R {\gv \sigma}   + {\gv {\tilde b}}{\mathbb A}_R {\gv \sigma} , \label{CMR}  
\end{align}
where R, L denotes the chiralities $\chi =+,-$ respectively, as indicated previously. Note that in this Hamiltonian there are 32 {independent} parameters contained in $\chi {\gv {\tilde b}}$, $\chi b _{0}$, $ {\gv v}_\chi $ and $  {\mathbb A}_\chi$.

To accomplish this task we now focus on the fermionic system described by the extended Dirac operator $(\Gamma^\mu i \partial_\mu- M)$ with the specific {choices in Eq.} (\ref{GAMMAMU}), which account for 40 independent parameters. A convenient simplification in the construction of the Dirac-Hamiltonian is to set $\Gamma^0=\gamma^0$, which demands {$c ^{0}_{\phantom{0} \nu} =0$} and  $d ^{0}_{\phantom{0} \nu}=0$. This cuts down the number of independent parameters in the SME to the required $40-8=32 $ and yields the Hamiltonian 
\begin{align}
    {\cal H}=\gamma^0 \, \Gamma^k i \partial_k +\gamma^0 M. 
    \label{SMEHAM}
\end{align}
In the chiral representation of the gamma matrices, where $\sigma_0$ is the $2\times2$ unit matrix and $\sigma^i$ are the standard Pauli matrices, we have
\begin{align}
    \gamma ^{0}= \begin{pmatrix} 0 & \sigma_0 \\  \sigma_0 & 0 \end{pmatrix} , \qquad  \gamma ^{i} = \begin{pmatrix} 0 & \sigma ^{i} \\  - \sigma ^{i} & 0 \end{pmatrix} , \qquad \gamma ^{5} = \begin{pmatrix} -\sigma_0 & 0 \\ 
0 & \sigma_0 \end{pmatrix},
\end{align}
and ${\cal H}$ separates into  right $(R)$ and left $(L)$ contributions according to the chiral projectors (\ref{11}). Before  writing them explicitly it is convenient to introduce the additional parametrization
\begin{align}
{{\delta^\mu}_\nu} +   c ^{\mu} _{\phantom{\mu}\nu} - \chi d ^{\mu} _{\phantom{\mu}\nu} =(m _{\chi}) ^{\mu} _{\phantom{\mu}\nu} \equiv \begin{pmatrix} 1 & 0 \\ V ^{i} _{\chi} & (U _{\chi}) ^{i} _{\phantom{i} j} \\    \end{pmatrix}, \label{Convenciones_V513}
\end{align} 
which arises from the choice $\Gamma ^{0} = \gamma ^{0}$. A direct calculation shows 
\begin{align}
    {\cal H} _{L}  &=  - V _{L} ^{i} k _{i} + k _{i} \,(U _{L})^{i} _{\phantom{i}j}\, \sigma ^{j} + C _{L0} - C _{Lj} \sigma ^{j} , \label{HAMSME_L} \\ {\cal H}_{R} &= -V _{R} ^{i} k _{i} - k _{i} (U_{R}) ^{i}_{\phantom{i}j} \sigma ^{j} + C _{R0} + C _{Rj} \sigma ^{j} . \label{HAMSME_R}
\end{align}
For further clarity, let us recap the steps we have followed:  we start with the parameters $ c ^{\mu}_{ \phantom{\mu} \nu}$, $d ^{\mu}_{ \phantom{\mu} \nu}$, $a _{\mu}$, $b _{\mu}$ in the SME, then the chiral representation naturally introduces $(m _{\chi}) ^{\mu}_{ \phantom{\mu} \nu}$ and $ (C _{\chi}) _{\nu}$ according to  (\ref{GAMMACHI}) and (\ref{141}). To fulfill the restriction $\Gamma ^{0} = \gamma ^{0}$ we have further introduced $ V ^{i} _{\chi}$ and $ (U _{\chi}) ^{i}_{ \phantom{i} j}$ in $(m _{\chi}) ^{\mu}_{ \phantom{\mu} \nu}$  as defined in Eq. (\ref{Convenciones_V513}). Summarizing, at this stage the SME parametrization of the Hamiltonian ${\cal H}$ is presented in terms of $V ^{i} _{\chi}$, $(U _{\chi}) ^{i} _{\phantom{i}j}$, ${(C _{\chi})} _{\mu}$, with $ \chi = \pm 1$.   

The final step is to write  the parameters of the SME Hamiltonian (\ref{SMEHAM}) in terms of those in  the  condensed matter Hamiltonian (\ref{Hamiltonian_Tilting}), i.e. to identify Eq. (\ref{CML}) with Eq. (\ref{HAMSME_L}) and Eq. (\ref{CMR}) with Eq. (\ref{HAMSME_R}). We obtain  the following relations 
\begin{align}
    & (m _{\chi}) ^{i} _{\phantom{i}0} = {V} ^{i} _{\chi} = {v} ^{i} _{\chi} , \qquad \qquad  \qquad (m _{\chi}) ^{i} _{\phantom{i}j} =  (U _{\chi}) ^{i} _{\phantom{i}j} = (A _{\chi}) _{ij} ,  \notag \\ & (C _{\chi}) _{0} = \chi \,( { v} ^{i} _{\chi} \,  {\tilde b} ^{i} - {\tilde b} _{0}), \qquad \hspace{1cm} (C _{\chi}) _{j} = \chi \, {\tilde b} ^{i} (A _{\chi}) _{ij} , \notag \\ &a _{0} = \frac{1}{2}{\tilde b} ^{i} ({v} ^{i} _{R} - {v} ^{i} _{L} ), \qquad \hspace{1.6cm} b _{0} = {\tilde b} _{0} - \frac{1}{2} {\tilde b} ^{i} ({v} ^{i} _{R} + {v} ^{i} _{L} ) , \notag \\ & a _{j} = \frac{1}{2} {\tilde b} ^{i} [ (A _{R}) _{ij} - (A _{L} ) _{ij} ] , \qquad  \hspace{0.3cm} b _{j} = {-} \frac{1}{2} {\tilde b} ^{i} [ (A _{R})_{ij} +(A _{L}) _{ij} ] .  \label{SUMMARY}
\end{align}

}

\section{The finite density regime }

\label{IV}
In order to apply the results of our effective action calculation {in Section \ref{II} }to obtain the conduction current produced by the WSM  characterized by the Hamiltonian (\ref{Hamiltonian_Tilting}) in the  zero temperature limit, we have to incorporate the effects of the chemical potential $\mu$ because  its location will determine the filling of the conduction and the valence bands of each node, thus yielding the conductivity. To this end we choose the Matsubara imaginary time formalism to 
to correctly incorporate the $\mu$-dependence in the effective action \cite{le_bellac_1996,Kapusta:2006pm}, which we implement through the substitution \cite{Bernard:1974bq,Dittrich:1979ux,Das:1997gg} 
\begin{align}
\int\frac{d ^{4} k}{(2 \pi ) ^{4} } \ \  \rightarrow \ \ 
{\Big(\int\frac{d ^{4} k}{(2 \pi ) ^{4} } \, \Big )_{T=0, \, \mu=0}} +i T \sum_ {n = - \infty} ^{\infty} \int \frac{d ^{3} {\gv k} }{(2 \pi ) ^{3}} , \label{25} 
\end{align}
in Eq. (\ref{24}). Here $k_{0}\to k_{0}=i\omega_{n}+\mu$ and  
the sum in Eq. (\ref{25}) is over the  Matsubara frequencies $\omega _{n} = (2n+1) \pi T$ required to produce antiperiodic boundary conditions for the fermions \cite{Kapusta:2006pm}. We think of the prescription (\ref{25}) as a natural regulator for potentially divergent contributions.

Since we are dealing with chiral fermionic excitations whose dispersion behaves linearly around the Weyl
nodes we have to locate the chemical potential close to the nodes in order for the approximation to be valid.
Once we do this, the 
calculation will tell us which band (valence or conduction) in each node contributes to the conduction process. Let us recall  that completely filled bands do not contribute to the current, which is due only to the partially filled bands. 

Next we focus on Eq. (\ref{24}) and  make use of the following relation  to evaluate the contribution of the  chemical potential {at zero temperature}
\cite{le_bellac_1996}  
 \begin{align}
 \lim _{T \to 0} T \sum _{n = - \infty} ^{\infty} g^\chi _{\rho} ( k _{0} = i \omega _{n} + {\mu} , {\gv k} ) & = \sum_{\rm {Re} \left( k^{\chi \#}_0 \right) < \mu} \rm{Res}\, \left[g^\chi_\rho(k_0, \mbf{k})\right]
 \label{Matsubara_fermion}
\end{align}
where $ k^{\chi \#}_0 $ {stand for the location} of the poles in $k_0$ of $g_\rho(k_0, \mbf{k})$, with ${\rm Res}$ denoting the corresponding residue.  The sum is made over all the existing  poles {having a real part less than the chemical potential }. 

Next we {split  $I_\rho^{\chi}(C)=I_\rho^{\chi (1)}(C) + I_\rho^{\chi (2)}(C)$, where the superindex $(1)$ refers to the $\mu$-independent contribution, while the superindex $(2)$ labels the  
 $\mu$-dependent piece.}
In turn, this says that  ${\cal B}_\mu={\cal B}^{(1)}_\mu+{\cal B}^{(2)}_\mu$

\subsection{The $\mu$-independent contribution}

\label{IVA}

This term reduces to  the standard  zero-temperature, zero-chemical potential piece, which has been previously obtained in the literature \cite{Zyuzin:2012tv, Goswami:2012db}. Going back to Eq. (\ref{24}), this corresponds to the direct evaluation of the integral $I_\rho(C)$
after the change of integration variables $k ^{\prime} _{\mu} = k _{\nu} m ^{\nu}{}_{\mu}$. Since the only vector at our disposal is $C_\mu$, we have 
\beq
I^{\chi (1)}_\rho(C_\chi)= \frac{i}{(\det m_\chi)}\,  N_\chi  (C_\chi)_\rho, \quad N_\chi=\frac{1}{C_\chi^2}\left[ \int \frac{d ^{4} k ^{\prime}}{\left( 2 \pi \right) ^{4}} \; \frac{ \left( k ^{\prime} - C_\chi \right) \cdot C_\chi}{\left[\left( k ^{\prime} - C_\chi \right) ^{2} \right] ^{2} } \right]_E , \label{26}
\eeq
where the integral inside the square brackets is in  Euclidean space and the factor $+i$ comes from the Wick rotation.
The factor $N_\chi$ is regularization dependent and could only be a function of the magnitude of the four-vector 
$(C_\chi)_{\mu}$. However, a change of scale $C_{\sigma} \to \lambda C _{\sigma}$ followed by an additional change of variables  $k ^{\prime \prime} _{\mu} = \lambda k ^{\prime} _{\mu}$ shows that $N_\chi$ is just a numerical factor, independent of $(C_\chi)_{\mu}$. Therefore, $N_\chi$ is the same for both left- and right-handed excitations, i.e. $N _{\chi}= N$. In this way, the total contribution to the vacuum polarization in this case is summarized in the vector
\begin{align}
{\cal B}^{(1)} _{\lambda} = 4 \pi ^{2} N \, \sum_{\chi= \pm 1} \chi \,  (C_\chi)_\rho \, (m_\chi^{-1}) ^{\rho} {}_{\lambda}, \label{29a} 
\end{align}
according to Eq. (\ref{BLAMBDA}). As shown previously in the literature the factor $N$ is finite but undetermined  \cite{Jackiw:1999yp, PhysRevLett.83.2518, Battistel:2000ms, Perez-Victoria:2001csb,Andrianov:2001zj,Altschul:2003ce,Altschul:2004gs, Alfaro:2009mr,Jackiw:1999qq,Altschul:2019eip}. 
 Its dependence  upon the regularization procedure has been studied in Ref. \cite{Goswami:2012db} and the final choice is  made  by selecting an observable quantity predicted by the model. In our case we take the anomalous Hall conductivity $\sigma_{xy}=-e^2 {\tilde b}_z/(2 \pi)^2$ as the quantity to be predicted in the isotropic zero-tilting limit, which results by selecting   $N=-1/(8\pi^2)$  \cite{Goswami:2012db}.
The method that we  employ to regularize the integral (\ref{26}), yielding the chosen value for the coefficient $N$, requires  to take a cutoff in the direction of $ (C_\chi)_\rho $ and involves performing a direct integration in cylindrical coordinates. Without loss of generality, we assume that the vector ${ {\boldsymbol{C}} _\chi}$ has only a component along the $z$-axis, such that we can conveniently split  $\mathbf{k}'=\mathbf{k'}_z+\mathbf{k'}_{\perp}$.
{As suggested in Ref.  \cite{Andrianov:2001zj},  we rely on the residual rotational symmetry  in the plane perpendicular to $\mathbf{C_\chi}$ to  perform first the integrations over $k'_\perp$ and $\phi$ }. Thus, the integral (\ref{26}) turns out to be
\begin{equation}
	N=\frac{1}{C_\chi^2(2\pi)^4} \int_{-\infty}^{\infty} dk'_0  \int_{-\infty}^{\infty} dk'_z  \int_0^{2\pi} d\phi  \int_{0}^{\infty} dk'_{\perp} k'_{\perp} \frac{(k'_0-(C_\chi)_0) (C_\chi)_0+ (k'_z-(C_\chi)_z)(C_\chi)_z}{((k'_0-{(C_\chi)_0})^2+{k'_{\perp}}^2+(k'_z-(C_\chi)_z)^2)^2}.\label{Nint}
\end{equation}
After integrating {the remainig variables} (see {the Appendix} \ref{APint}), we find that $N=-1/(8\pi^2)$ as agreed.

\subsection{The $\mu$-dependent contribution} 

\label{IVB}

{Using the result (\ref{Matsubara_fermion}), we next consider the finite density contribution to Eq. (\ref{24}), and compute}
\begin{equation}
I^{\chi(2)}_{\rho }=i \int 
\frac{d^{3}\mathbf{k}}{(2\pi )^{3}}\left[ \sum_{\rm {Re} ( k^{\chi \#}_0) \,  < \mu} {\rm{Res}}\left[g^\chi_\rho(k_0, \gv{k})\right] \right]. \label{Integral_mu_dependent}
\end{equation}%
The calculation of the poles and residues is summarized in the Appendix \ref{APC}. The poles for the  band $s$  are located at
\begin{align}
 k^{\chi \#} _{0 s}= - k _{j} (m_\chi) ^{j}{}_{0} + (C_\chi) _{0} + s \,  \vert {\gv k} ^{\prime} - {\gv C_\chi}  \vert , \quad s = \pm 1 ,  \quad k ^{\prime} _{j} = k _{i} \,(m_\chi) ^{i}{}_{j}.
\label{POLES}
\end{align}
It is a simple matter to verify that these poles correspond to the dispersion relations (\ref{Energy}) of the WSM Hamiltonian
(\ref{Hamiltonian_Tilting}), as expected.

The residues of $g^\chi_\rho(k_0, \mbf{k})$ at the poles $k^{\chi \#} _{0 s}$  are the real expressions
\begin{equation}
\mathrm{Res} [g^\chi_{\rho}( k _{0}, \gv{k})] =  -s \frac{1}{4} \delta ^{j} _{\rho} \frac{k ^{\prime}_j - (C_\chi)_{j}}{ \vert {\gv k} ^{\prime} - {\gv C}_\chi \vert ^{3} } . \label{Residues}
\end{equation}
Then we have to evaluate
\barr
I^{\chi (2)}_{\rho }&=&-i \delta^j_\rho   \sum_{ s =\pm 1} s \int 
\frac{d^{3}\mathbf{k}}{(2\pi )^{3}} \frac{k ^{\prime}_j - (C_\chi)_{j}}{ \vert {\gv k} ^{\prime} - {\gv C}_\chi \vert ^{3} } H(\mu-k_{0 s}^{\chi \#})\nonumber \\
&=& -i \delta^j_\rho  \frac{1}{\det (m_\chi)} \sum_{ s=\pm 1} s \int 
\frac{d^{3}\mathbf{k''}}{(2\pi )^{3}} \frac{k''_{j} }{ \vert {\gv k}'' \vert ^{3} } H(\mu-k_{0 s}^{\chi \#}),
\earr
where we introduce  the convenient change of variables 
\begin{align}
k ^{\prime \prime} _{j} = k ^{\prime} _{j} - (C_\chi)_{j} , \quad d ^{3} {\gv k} ^{\prime \prime} = d ^{3} {\gv k} ^{\prime} = \det( m_\chi ) \, d ^{3} {\gv k}
\label{CHANGEV}
\end{align}
and  use $\det ((m_\chi) ^{i}{} _{j}) = \det ((m_\chi) ^{\mu}{}_{\nu} )  =\det (m_\chi)$, since $(m_\chi)^{0}{}_{\nu} = \delta ^{0} _{\nu}$.

The next step is to calculate
\beq
I_j^{\chi s}=\int 
\frac{d^{3}\gv{k}''}{(2\pi )^{3}} \frac{k''_{j} }{ \vert {\gv k}'' \vert ^{3} } H(\mu-k_{0 s}^\#).
\label{34}
\eeq
Recalling Eq. (\ref{23}), at this stage we can write the axial part of the  vacuum polarization as 
\begin{align}
{ \Pi} ^{\mu \nu} _{A, \, \chi } (p) = i \chi \frac{e ^{2}}{2} 
(m_\chi^{-1}) ^{j}{}_{\lambda} \, p _{\kappa} \, \epsilon ^{\mu \nu \lambda \kappa}\, \sum_{s= \pm 1} s I _{j} ^{\chi s}.
\label{VPSEMIF}
\end{align}
Let us now rewrite  the  poles $ k^{\chi \#} _{0 s}$,  in terms of the  new double-primed variables $ k ^{\prime \prime} $ which enter in the integral (\ref{34}). Starting from (\ref{POLES}), the sequence $k_i (m_\chi)^i{}_0=k'_j(m_\chi^{-1})^j{}_i\, (m_\chi)^i{}_0=(k''_j+(C_\chi)_j)
(m_\chi^{-1})^j{}_i\, (m_\chi)^i{}_0 $ yields 
\begin{align}
 k^{\chi \#} _{0 s}= \gv{\mathcal{V}}_\chi \cdot {\gv k} ^{\prime \prime}  + s \, \vert {\gv k} ^{\prime \prime}  \vert + E_{\chi 0}, \label{Condition}
\end{align}
with the additional definitions
\beq
{\cal V}_\chi^j=(m_\chi^{-1})^j{}_i\, (m_\chi)^i{}_0, \qquad E_{\chi 0}={\gv{\cal V}}_\chi \cdot {\gv{C}}_\chi + (C_\chi)_0
\label{ADDDEF}
\eeq
 A proof of consistency can be made recalling that  
$k^{\chi \#} _{0 s} $ describe  the energy bands of the WSM. In the notation of (\ref{Condition}) we see that the nodes are located at $\gv{k}''=0$, such that the energy of each  node has to be  $E_{\chi 0}$, which we know it is equal to $-\chi {\tilde b}_0$. This result is recovered from the alternative expression for $E_{\chi 0}$ in Eq.(\ref{ADDDEF}) after the relations (\ref{SUMMARY}) are used. 

The simplest way to calculate $I_j^{\chi s}=-\left( I^{\chi s} \right)^j$ is to realize that 
$\gv{k}/|\gv{k}|^3
=-{\gv \nabla}(1/|\gv{k}|)$, which calls for an integration by parts yielding a surface term plus a volume contribution involving $\delta(k_{0 s}^*- \mu)$. Also it is convenient   
to use spherical coordinates choosing the  $z$-axis in the direction of $\gv{\cal V}_\chi$, such that 
\beq
k^{\chi \#} _{0 s} = | {\gv k} ''| \, \big(|\gv{\mathcal{V}}_\chi| \cos \theta + s \big)+E_{\chi 0}.
\label{KSPHE}
\eeq 
In the following we restrict ourselves to Type I WSMs where $|\gv{\cal V}_\chi| < 1$, implying that the sign of the factor of $|{\gv{k}}''|$ in (\ref{KSPHE}) is independent of the angle $\theta$, i. e. we have that $(1+s |\gv{\mathcal{V}}| \cos \theta)$ is always positive. The detailed evaluation of the integral (\ref{34}) is outlined in Appendix \ref{APD} and we only present the results here. When $|\gv{\cal V}| < 1$ the surface integral vanishes and we are left with 
\beq
\{ \left( I^{\chi s} \right)^j \} ={\gv I}^{\chi s}=\int 
\frac{d^{3}\gv{k}''}{(2\pi )^{3}} \frac{1}{ \vert {\gv k}'' \vert } \gv{\nabla} H(\mu-k_{0 s}^*).
\label{341}
\eeq
The resulting delta function imposes the condition $k^{\chi \#} _{0 s} =\mu$ which we  write as 
\beq
| {\gv k} ''| \, \big(|\gv{\mathcal{V}}_\chi| \cos \theta + s \big)=
\mu+\chi {\tilde b}_0 \equiv \Lambda_\chi
\label{DEFLAMBDA}
\eeq
Choosing $\gv{W}=\gv{\mathcal{V}}_\chi$ in Eq. (\ref{Integral_I_appendix_2}) of the Appendix \ref{APD} we find
\beq
{\gv I}^{\chi s}= \Big( \frac{1}{2 \pi^2}\frac{1}{|{\gv {\cal V}}_\chi|^3}  (s\, \Lambda_\chi) H(s \Lambda_\chi)\Big[|{\gv {\cal V}}_\chi|-{\rm arctanh} |{\gv {\cal V}}_\chi| \Big] \Big)  {\gv {\cal V}}_\chi,
\eeq
which demands $(s \, \Lambda_\chi)$ to be positive. Going  back to the vacuum polarization tensor (\ref{VPSEMIF}) we consider a given Weyl node (fixed $\chi$) and evaluate its contribution. Let us examine the location of the chemical potential $\mu$ relative to the position of the energy of node $E_{\chi 0}=-\chi {\tilde b}_0 $. When $\mu$ is above $E_{\chi 0}$ we have $\Lambda_\chi >0$ in such a way that only the conduction band $s=1$ contributes with 
\beq
[s {\gv I}^{\chi s}]_{s=+1}= \frac{1}{2 \pi^2}\frac{1}{|{\gv {\cal V}}_\chi|^3} \Lambda_\chi \Big[|{\gv {\cal V}}_\chi|-{\rm arctanh} |{\gv {\cal V}}_\chi| \Big] {\gv {\cal V}}_\chi.
\eeq
When $\mu$ is below $E_{\chi 0}$ we have the opposite situation where $\Lambda_\chi$ is negative and only the valence band $s=-1$
contributes with
\beq
[s {\gv I}^{\chi s}]_{s=-1}=\frac{1}{2 \pi^2}\frac{1}{|{\gv {\cal V}}|^3} \Lambda_\chi \Big[|{\gv {\cal V}}_\chi|-{\rm arctanh} |{\gv {\cal V}}_\chi| \Big] {\gv {\cal V}}_\chi.
\eeq
Summarizing, when $|{\gv {\cal V}}| < 1$ 
{the contribution of each node to the current results proportional to the corresponding value of  $\Lambda_\chi$, independently of which band provides the conducting charges,  since Eq. (\ref{VPSEMIF}) requires to consider the product}
$s {\gv I}^{\chi s}$ in each case. In other words, $\pi^2 [s {\gv I}^{\chi s}]_{s=+1}=\pi^2 [s {\gv I}^{\chi s}]_{s=-1}\equiv N_\chi {\gv {\cal V}}_\chi$ with
\beq
N_\chi=\frac{1}{2 |{\gv {\cal V}}|^3} \Big[|{\gv {\cal V}}_\chi|-{\rm arctanh} |{\gv {\cal V}}_\chi| \Big].
\label{NCHI}
\eeq
The final result  for Eq.  (\ref{VPSEMIF}) is
\begin{align}
{ \Pi} ^{\mu \nu} _{A, \, \chi } (p) = i \frac{e ^{2}}{2 \pi^2} 
\, p _{\kappa} \, \epsilon ^{\mu \nu \lambda \kappa}\,\sum_{\chi=\pm 1} \chi \Lambda_\chi  N_\chi  (m_\chi^{-1}) ^{j}{}_{\lambda}           ({\cal V}_\chi)_j,
\end{align} 
from where we read 
\beq
{\cal B}^{(2)}_\lambda=-\sum_{\chi=\pm 1} \chi \Lambda_\chi  N_\chi  (m_\chi^{-1}) ^{j}{}_{\lambda} ({\cal V}_\chi)_j,
\label{BLAMBDA2}
\eeq
according to (\ref{BLAMBDA}).
In the Appendix \ref{APE} we {put together the separate contributions to ${\cal B}_\lambda$ in Eqs. (\ref{29a}) and (\ref{BLAMBDA2}), expressing the final result in terms 
 of the parameters of the WSM Hamiltonian (\ref{Hamiltonian_Tilting}). We obtain}
\barr
&&{\cal B}_0= {\tilde b}_0 -\sum_{\chi= \pm 1} \chi \Lambda_\chi \, N_\chi \, {v}^i_\chi \big[({\mathbb A}_\chi^{-1})^T  ({\mathbb A}_\chi^{-1})\big]_{ij} {v}^j_\chi, \label{GENREL0} \\
&&{\cal B}_k= {\tilde b}_k + \sum_{\chi=\pm 1} \chi \Lambda_\chi \, N_\chi \, 
\left[({\mathbb A}_\chi^{-1})^T ({\mathbb A}_\chi^{-1})
\right]_{k l} \, {v}^l_\chi, 
\label{GENREL1}
\earr
with
\begin{equation}
 \Lambda_\chi= \mu + \chi {\tilde b}_0, \qquad 
({\cal V}_\chi)^i= ({\mathbb A}^{-1}_\chi)_{ik} {v}^k_\chi, \qquad N_\chi=\frac{1}{2 |{\gv{\cal V}}_\chi|^3} \left( |{\gv{\cal V}}_\chi|-{\rm arctanh}(|{\gv{\cal V}}_\chi|) \right), \qquad |{\gv{\cal V}}_\chi|=\sqrt{({\cal V}_\chi)^i({\cal V}_\chi)^i}.
\label{GENREL11}
\end{equation}

\section{A chiral kinetic theory approach} \label{Kinetic_Theory_App}

We now validate our results by using chiral kinetic theory, which is a topologically modified semiclassical Boltzmann formalism to describe the behavior of Weyl fermions at finite chemical potential. In the presence of an electric field, in addition to the usual band dispersion, the velocity for Bloch electrons acquires an extra term proportional to the Berry curvature \cite{RevModPhys.82.1959, PhysRevLett.99.236809}. This gives rise to a transverse topological current given by 
\begin{align}
{\bf{J}} = - \frac{e ^{2}}{\hbar } \sum _{s} \sum _{\chi = \pm 1} \int \frac{d ^{3} {\boldsymbol{k}} }{(2 \pi ) ^{3}} {\bf{E}} \times \boldsymbol{\Omega} _{s \chi} ({\boldsymbol{k}}) \, f _{s \chi} ^{\mbox{\scriptsize F.D.}} ({\boldsymbol{k}}) ,  \label{CurrentDensity}
\end{align}
where ${\bf{E}}$ denotes the electric field, $ f _{s \chi} ^{\mbox{\scriptsize F.D.}} ({\boldsymbol{k}}) $ is the Fermi-Dirac distribution function for Bloch electrons with chirality $\chi$ in the $s$-th band and $\boldsymbol{\Omega} _{s \chi } ({\boldsymbol{k}}) = i \bra{  \nabla _{{\boldsymbol{k}}}  u _{s \chi} ({\boldsymbol{k}}) } \times  \ket{  \nabla _{{\boldsymbol{k}}} u _{s \chi} ({\boldsymbol{k}}) } $ is the Berry curvature.  Here, the Bloch states $\ket{u _{s \chi} ({\boldsymbol{k}}) } $ are defined by $\hat{H} _{\chi} \ket{u _{s \chi} ({\boldsymbol{k}}) } = E _{s \chi} \ket{u _{s \chi} ({\boldsymbol{k}}) }$, where $\hat{H} _{\chi}$ is the single particle Hamiltonian for a Weyl fermion with chirality $\chi$. Evaluation of the Hall current from Eq. (\ref{CurrentDensity}) reproduces the Karplus-Luttinger formula for the anomalous Hall conductivity \cite{PhysRev.95.1154}.  In the following we evaluate the anomalous Hall current (\ref{CurrentDensity}) for a WSM described by the model Hamiltonian (\ref{Hamiltonian_Tilting}) characterized by the tilting ${\gv v} _{\chi}$ and the matrix of the Fermi velocities ${\mathbb A} _{\chi}$, which we consider symmetric in this section.

The topological properties of the two-node model $H _{\chi} (\gv{k}) $ under consideration can be seen from the Berry curvature. Using the Bloch states $\ket{u _{s \chi} ({\boldsymbol{k}}) } $ the Berry curvature is found to be
\begin{align}
\boldsymbol{\Omega} _{s \chi}  (\gv{k}) = - \frac{s \chi}{2} \det ({\mathbb A} _{\chi}) \frac{ {\mathbb A} _{\chi} ^{-1} {\boldsymbol{\mathcal{K}}} _{\chi} }{\mathcal{K} _{\chi} ^{3}} .  \label{BerryCurvature}
\end{align}
In the isotropic limit this expression reduces to the usual monopole-like Berry curvature of the Weyl nodes. Furthermore, we can check that the Berry flux piercing any surface enclosing the node is exactly $2 \pi \chi$. From Eq. (\ref{BerryCurvature}) we compute the topological current. First, we rewrite Eq. (\ref{CurrentDensity}) in the standard form of the anomalous Hall current ${\bf{J}} = \frac{e ^{2}}{2 \pi ^{2} } \delta \boldsymbol{\mathcal{B}} \times {\bf{E}}$, where
\begin{align}
\delta \boldsymbol{\mathcal{B}} =  2 \pi ^{2} \sum _{s} \sum _{\chi = \pm 1} \int \frac{d ^{3} {\boldsymbol{k}} }{(2 \pi ) ^{3}} \boldsymbol{\Omega} _{s \chi} ({\boldsymbol{k}}) \, f _{s \chi} ^{\mbox{\scriptsize F.D.}} ({\boldsymbol{k}}) .  \label{B-vector}
\end{align}
{Here we have introduced $\delta \boldsymbol{\mathcal{B}}$ instead of the full $\boldsymbol{\mathcal{B}}$ of the previous section because the semiclassical approximation fails to predict the Hall current proportional to the Weyl node separation. This is expected because the semiclassical approximation accounts for the single-band Fermi surface properties of the wave packets, and the Hall conductivity carries information of all filled states.} To evaluate the integral (\ref{B-vector}) we change the integration variable from $\gv{k}$ to ${\boldsymbol{\mathcal{K}}} _{\chi}$.  Substituting the Berry curvature (\ref{BerryCurvature}), the expression (\ref{B-vector}) in the zero-temperature limit becomes
\begin{align}
\delta \boldsymbol{\mathcal{B}} =  - \pi ^{2} \sum _{s} \sum _{\chi = \pm 1}  s \chi  {\mathbb A} _{\chi} ^{-1}   \int \frac{d ^{3} {\boldsymbol{\mathcal{K}}} _{\chi} }{(2 \pi ) ^{3}}   \frac{ {\boldsymbol{\mathcal{K}}} _{\chi} }{\mathcal{K} _{\chi} ^{3}} \, H \left( \mu + \chi {\tilde b}_{0} - {\boldsymbol{\mathcal{V}}} _{\chi} \cdot {\boldsymbol{\mathcal{K}}} _{\chi} - s \mathcal{K} _{\chi} \right) . \label{B-vector-2}
\end{align}
This integral is exactly the same obtained when evaluating the $\mu$-dependent contribution to the vacuum polarization tensor in Section \ref{IVB}. See Eq. (\ref{34}) for instance. A detailed solution of this integral is presented in the Appendix \ref{APD} and here we take the final result only.

For a type-I WSM, defined by $\mathcal{V} _{\chi} = \vert {\boldsymbol{\mathcal{V}}} _{\chi} \vert < 1$, the $\mu$-dependent correction to the anomalous Hall current is expressed in terms of the vector
\begin{align}
\delta \boldsymbol{\mathcal{B}} =  -  \sum _{\chi = \pm 1} \frac{\chi \Lambda _{\chi}}{2 \mathcal{V} _{\chi} ^{3}} \left[ \mathcal{V} _{\chi} - \mbox{arctanh} (\mathcal{V} _{\chi}) \right] \, {\mathbb A} _{\chi} ^{-1} {\mathbb A} _{\chi} ^{-1} {\gv v} _{\chi}  , \label{B-vector-fin}
\end{align}
where $\Lambda _{\chi} = \mu + \chi {\tilde b}_{0}$. This result reproduces the one obtained in the previous section, where it was derived within a quantum field theoretical approach, since $\delta{\gv{\cal B}}$ is equal to ${\gv{\cal B}}^{(2)}$ defined in the previous section.

\section{Applications}

\label{APPL}

In Weyl semimetals  the energy spectrum around band-touching points behaves as tilted and anisotropic Dirac cones. These differences from the ideal Dirac cone picture, where the charge carriers  behave as massless relativistic particles, have important consequences for the physical properties of Weyl semimetals as well as for their  optical and transport features.

In recent years, there has been a great deal of interest in measuring transport properties induced by the Berry curvature, specially the anomalous Hall effect \cite{RevModPhys.82.1539}.  In experiments, there are a number of factors which make difficult the measurement of the ideal anomalous Hall effect ${\bf{J}} = \frac{e ^{2}}{2 \pi ^{2} } {\boldsymbol{{\tilde b}} }  \times {\bf{E}}$.  For example, since the Weyl nodes do not lie exactly at the Fermi level, other factors are expected to be relevant, such as the anisotropy, tilting and disorder.  However, as shown in Ref. \cite{PhysRevLett.113.187202}, disorder-induced contributions to the anomalous Hall conductivity are absent when the Fermi level is near the nodes, thus leaving the anisotropy and tilting as possible responsible of deviations from the ideal prediction. In this Section we aim to fill in this gap. Here, we use the theory developed in the previous sections to investigate the anomalous Hall current for different WSM materials, mainly focusing on the effects of the tilting and the anisotropy.

\subsection{ The ideal Weyl semimetal  $\mbox{EuCd}_{2}\mbox{As}_{2}$}

\label{APPLA}

The simple structure of Weyl nodes in the trigonal crystal $\mbox{EuCd}_{2}\mbox{As}_{2}$ makes it an ideal material with which to study the different contributions to the anomalous Hall effect.  \textit{Ab initio} electronic structure calculations reveal that $\mbox{EuCd}_{2}\mbox{As}_{2}$ has only a single pair of Weyl nodes located at $\pm {\boldsymbol{{\tilde b}}} = (0,0, \pm \tilde{b} _{z})$ along the $c$ axis,  with ${{\tilde b}} _{z} = 0.03 \times 2 \pi / c \approx 0.26 \mbox{ nm} ^{-1}$.  Inversion symmetry guarantees that the nodes lie at the same energy, i.e. $\tilde{b} _{0} = 0$. If the nodes lie at the Fermi level,  \textit{ab initio} calculations predict the anomalous Hall conductivity to be $\sigma _{yx} ^{\mbox{\scriptsize AHC}} = \frac{e ^{2} \tilde{b} _{z} }{2 \pi ^{2} \hbar} \approx 30 \, \Omega ^{-1} \mbox{cm}^{-1}$, which is significantly larger than the observed value which is of the order of $0.5 \, \Omega ^{-1} \mbox{cm}^{-1}$ \cite{PhysRevB.100.201102}.  This prediction is valid when the nodes lie at the Fermi level, however, in the experiments carried out by authors in Ref.  \cite{PhysRevB.100.201102}, the nodes are slightly shifted from the Fermi level. In this case, as our results in Eqs. (\ref{GENREL1}) and (\ref{B-vector-fin}) anticipate, the anisotropy and tilting contribute to the anomalous Hall conductivity by replacing {$\boldsymbol{{\tilde b}}$} to ${\boldsymbol{{\tilde b}}} + \delta \boldsymbol{\mathcal{B}}$. In the following we estimate such corrections. To this end, we use the low-energy linearly-dispersing two-band model derived in Ref. \cite{PhysRevB.100.201102}. The effective Hamiltonian for a Weyl node of chirality $\chi$ is given by Eq. (\ref{Hamiltonian_Tilting}), with ${\gv v} _{\chi} = (0,0,\chi v )$ and ${\mathbb A} _{\chi} = \chi \, \mbox{diag} (v _{\parallel} , v _{\parallel} , \chi v _{\perp})$. These velocity parameters are obtained by fitting the energy computed from \textit{ab initio} calculations.  The obtained values are: $v = 1.6 v _{0}$, $v _{\parallel} = 3 v _{0}$ and $v _{\perp} = 2 v _{0}$, where $v _{0}=1.51 \times 10 ^{5}$ m/s. Furthermore, from Shubnikov-de Haas measurements and \textit{ab initio} calculations, the Fermi level is predicted to be approximately $52$ meV below the Weyl nodes thus lying in the valence band. All in all, the shift of the Weyl node position (\ref{B-vector-fin}) is found to be $\delta \boldsymbol{\mathcal{B}} = (0, 0 , - 0.122) \mbox{ nm}^{-1}$, such that the anomalous Hall conductivity becomes
\begin{align}
\tilde{\sigma } _{yx} ^{\mbox{\scriptsize AHC}} = \frac{e ^{2} ({{\tilde b}} _{z} + \delta \mathcal{B} _{z} )}{4 \pi ^{2} \hbar} \approx 15.9 \, \Omega ^{-1} \mbox{cm}^{-1} , 
\end{align}
which represents a significant reduction from the originally predicted anomalous Hall conductivity $\sigma _{yx} ^{\mbox{\scriptsize AHC}}$. However there are other factors which could diminish further the conductivity, for example, finite temperature effects and higher-order terms in the model Hamiltonian.

\subsection{ The highly anisotropic and tilted monopnictide TaAs}

\label{APPLB}

We will now apply our results to the archetypal non-magnetic transition-metal monopnictide TaAs, the best-known WSM \cite{PhysRevX.5.031013}. This material crystallizes in a body-centered tetragonal structure with lattice parameters $a = b = 0.34348 \mbox{ nm} ^{-1}$ and $c = 1.1641 \mbox{ nm} ^{-1}$. In the Brillouin zone of TaAs, there are 24 Weyl nodes in total: 8 Weyl nodes on the $k_{z}= 2 \pi /c$ plane (W1 nodes) and 16 Weyl nodes away from the $k_{z}= 2 \pi /c$ plane (W2 nodes).  W1 (W2) nodes are located $26$ meV ($13$ meV) below the Fermi level, which crosses only the conduction band (electron pockets).  The electronic structure of TaAs shows that the Weyl bands around the W1 and W2 nodes possess strong anisotropies and tilting. Since the Weyl nodes are separated in the $k _{x}$-direction, a linear fit of the Weyl bands near the nodes produces an effective Hamiltonian as given by Eq. (\ref{Hamiltonian_Tilting}),  such that the tilting and matrix of Fermi velocities depend if the node is either W1 or W2. Extracted from \textit{ab initio} calculations of Ref. \cite{10.1140/epjb/e2020-10110-x},  for W1 nodes the band parameters are
\begin{align}
\mathbb{A} _{\chi} ^{\mbox{\scriptsize W1}} = \chi v' _{0} \left( \begin{array}{ccc}  3.963 \, \chi &  0.393 \, \chi & 0 \\  0.393 \, \chi & 2.318 & 0 \\ 0 & 0 & 0.212 \end{array} \right) , \qquad   {\gv v} _{\chi} ^{\mbox{\scriptsize W1}} = v' _{0} \left( \begin{array}{c} - 1.603 \, \chi \\ 1.004 \\0 \end{array} \right) ,
\end{align}
and for the W2 nodes we have
\begin{align}
\mathbb{A} _{\chi} ^{\mbox{\scriptsize W2}} = \chi v' _{0} \left( \begin{array}{ccc}  3.220 \, \chi &  1.127 \, \chi & 0.661 \, \chi  \\  1.127 \, \chi & 0.291 & 2.464 \\ 0.661 \, \chi  & 2.464 & 1.659 \end{array} \right) , \qquad   {\gv v} _{\chi} ^{\mbox{\scriptsize W2}} = v' _{0} \left( \begin{array}{c} - 0.989 \, \chi \\ 0.944 \\ 1.409 \end{array} \right) ,
\end{align}
where $v' _{0} =1 \times 10^{5}$m/s. These parameter values support the picture of  highly anisotropic and strongly tilted Weyl cones in TaAs. Besides, note that W1 Weyl bands are almost 2D, while W2 bands are 3D. With all the information  above, we are able to compute the correction to the anomalous Hall conductivity by using our equivalent formulas (\ref{BLAMBDA2}) and  (\ref{B-vector-fin}). The contributions to the shifting from the W1 and W2 nodes are found to be 
\begin{align}
\delta \boldsymbol{\mathcal{B}} ^{\mbox{\scriptsize W1}} = \left( \begin{array}{c} 0.01719 \\ -0.00655 \\ 0 \end{array} \right)  \mbox{ nm}^{-1}  , \qquad  \delta \boldsymbol{\mathcal{B}} ^{\mbox{\scriptsize W2}} = \left( \begin{array}{c} -0.04814 \\ 0.02670 \\ 0.01967 \end{array} \right)  \mbox{ nm}^{-1} ,
\end{align}
respectively. Here, we used that $\Lambda ^{\mbox{\scriptsize W1}} _{\chi} = 26$ meV and $\Lambda ^{\mbox{\scriptsize W2}} _{\chi} = 13$ meV, which are the positions of the W1 and W2 nodes below the Fermi level.  We observe that the shifting of the W1 nodes are confined to the $x$-$y$ plane, consistent with the fact that Weyl bands are almost 2D. The full contribution to the anomalous Hall conductivity can be obtained by summing up the contributions from the 12 pairs of nodes of TaAs, i.e.
\begin{align}
\delta \boldsymbol{\mathcal{B}} = \sum _{i=1,2} N _{i} \, \delta \boldsymbol{\mathcal{B}} ^{\mbox{\scriptsize W} i} = \left( \begin{array}{c} -0.31639 \\ 0.18741 \\ 0.15743 \end{array} \right)  \mbox{ nm}^{-1} , 
\end{align}
where we used $N _{1} = 4$ and $N _{2} =8$, which are the number of pairs of $W1$ and $W2$ nodes in the Brillouin zone of TaAs. Interesting conclusions can be extracted from this result. First, we observe that the effects of the anisotropy and tilting can be clearly distinguished from the conventional anomalous Hall current, which is proportional to the Weyl nodes separation. In the present case the nodes are separated along the $x$ axis and hence the anomalous Hall conductivity is $\sigma _{zy}  ^{\mbox{\scriptsize AHC}} = \frac{e ^{2} {{\tilde b}}_{x} }{2 \pi ^{2} \hbar}$. However, in the presence of anisotropy and tilting, the resulting anomalous Hall current is
\begin{align}
\tilde{\sigma} _{zy}  ^{\mbox{\scriptsize AHC}} = \frac{e ^{2} ({{\tilde b}} _{x} + \delta \mathcal{B} _{x} ) }{2 \pi ^{2} \hbar} , \qquad \tilde{\sigma} _{yx}  ^{\mbox{\scriptsize AHC}} = \frac{e ^{2} \delta \mathcal{B} _{z} }{2 \pi ^{2} \hbar} , \qquad \tilde{\sigma} _{xz}  ^{\mbox{\scriptsize AHC}} = \frac{e ^{2} \delta \mathcal{B} _{y} }{2 \pi ^{2} \hbar} .
\end{align}
Therefore, the two last terms are purely induced by anisotropy and tilting.

Finally, it is worth mentioning that our analysis can also be applied to other WSM materials of the TaAs family, such as TaP, NbAs and NbP. These materials are also highly anisotropic and tilted, however,  unlike TaAs where only {electrons pockets} occurs,  \textit{ab initio} calculations indicate that in these other WSMs both {electron pockets} for W1 nodes and {hole pockets} for W2 nodes occur \cite{10.1140/epjb/e2020-10110-x}.

{\section{Summary and results}}

\label{VII}

{

In quantum field theory the calculation of the effective  action $S_{\rm eff}(A^\mu)$ starting from a fermionic system minimally coupled  to external  electromagnetic fields $A^\mu$  provides a very general method to obtain the  effective  current $J^{\rm eff}_\mu(x)=\delta S_{\rm eff}/\delta A^\mu(x)$  yielding the response of the system. For  a wide class of materials this procedure  gives an alternative 
to some strategies  frequently used in condensed matter physics to determine the electromagnetic response of a medium, such as   
to the Kubo linear response theory and  the chiral kinetic theory approach, for example. 
Such particular set of materials
include fermionic excitations having a  dispersion relation which is linear near the band-touching points in the Brillouin zone close to the Fermi energy, thus signaling the presence of Dirac-Weyl quasiparticles akin to the real particles  resulting in the fundamental interactions of high energy physics. Nevertheless, a fundamental difference arises since the periodic and bounded structure of a crystaline lattice induce the violation of the spacetime symmetries such as translations, continuous rotations and Lorentz invariance.
Fortunately, a quantum field theory model has been constructed to deal with such violations: the Standard Model Extension (SME). Its   minimal fermionic sector covers many Dirac-Weyl Hamiltonians coinciding with the linearized version, near the Fermi energy, of the tight-binding Hamiltonians describing materials such as topological insulators and Dirac-Weyl semimetals, for example.

Having in mind the application of our results to the realistic case of a Weyl semimetal with tilting and anisotropy  we considered  the selection (\ref{GAMMAMU}) of parameters in the SME, which embodies those included in the condensed matter Hamiltonian (\ref{Hamiltonian_Tilting}). The resulting fermionic action is chiral, which simplifies enormously the calculation of the CPT-odd contribution of the vacuum polarization tensor $\Pi^{\mu\nu}_{A}$ that can be further  split into the sum of the two chiral contributions. This tensor 
 determines the CPT-odd effective action in our approximation of two powers in the external electromagnetic field. We restrict ourselves to this sector since it produces novel effects such as the  anomalous Hall current and a contribution to the chiral magnetic effect. Our calculation is nonperturbative in the parameters of the model and yields the general result that the ensuing  effective electrodynamics remains of the axionic type  (\ref{02})  with $\Theta(x)= {\cal B}_\mu x^\mu$, where ${\cal B}_\mu$ is the main objective of the calculation being related to $\Pi^{\mu\nu}_{A}$ through Eq. (\ref{BLAMBDA}). The results, previous to the inclusion of finite density effects are given in Eq. (\ref{CHIRALB}) for each chirality. They are presented in terms of the chosen LIV parameters of the SME, but can also be expressed  using those of the condensed matter Hamiltonian employing Eqs. (\ref{SUMMARY}).  
 
 The filling of the valence and  the conduction bands with respect to the Fermi energy determines the  conduction properties of the material and we introduce this dependence through the chemical potential $\mu$. This naturally split the basic integral (\ref{CHIRALB}) into a $\mu$-independent piece plus a $\mu$-dependent one. In the first contribution we face the well-known and much discussed problem that the overall factor of the integral is finite but undetermined. We fix this ambiguity by selecting the  factor $N=-1/(8\pi)^2$ that reproduces the anomalous Hall conductivity $\sigma_{xy}=-e^2 {\tilde b}_z/(2\pi)^2$ in the isotropic case with no tilting and when the cones are separated along the $z$-axis.
 {After we rely on  the residual  rotational symmetry in the presence of the vector ${\mathbf C}$ to perform a first integration, we
  show that this factor is obtained when we introduce  a cutoff along the direction where   each of the remaining  integrals diverge.} To deal with the second $\mu$-dependent contribution we extend  the integral (\ref{24}) to the imaginary-time  using the Matsubara formalism  at zero temperature and   calculate it  following the  prescription  (\ref{Matsubara_fermion}). This yields our final result for the vector
 ${\cal B}_\lambda$ presented in Eqs. (\ref{GENREL0}), (\ref{GENREL1})  and (\ref{GENREL11}) in terms of the parameters of the condensed matter Hamiltonian.
 
 We have also obtained the anomalous Hall current from the chiral kinetic theory approach. In this case,  within the linear approximation we are considering, the predicted value for the $\mu$-independent part is zero. Assuming a symmetric anisotropy matrix we get the result in (\ref{B-vector-fin}) for the  $\mu$-dependent contribution, which coincides with that in Eq. (\ref{GENREL1}) calculated via the effective action. The contribution from ${\cal B}_0$ can also be obtained within the semiclassical approach. This agreement, together with the result itself, constitutes the most important conclusion of our work showing the strength of effective action calculations in WSMs. 
 
 Our final results are further limited to type-I WSMs, where the magnitude of the effective tilting parameter $\gv{\cal V}_\chi$ is less than one. In this case the integral ${\mathbf I}^{\chi s}$, common to both methods and calculated in the Appendix \ref{APD}, has no singularities in the angular range of integration. Different from type-I WSMs, the energy values of Weyl nodes in type-II WSM are not the local extrema. Their discussion would require introducing additional cutoffs in the model, which may be material dependent, so we consider this situation beyond the scope of the present work. In our case the contribution to the conductivity of each node is due to a single band $s$ which is determined by the sign of $\Lambda_\chi=\mu-E_0$ according to the condition $s\Lambda_\chi >0$.  The constant $\Lambda_\chi$ measures the location of the chemical potential with respect to the energy of the node $E_0$.
 
  We include two applications that show the important consequences of the tilted and anisotropic Dirac cones for the transport  properties of Weyl semimetals, focusing upon the anomalous Hall current.
  
  In the first case of the ideal Weyl semimetal $\mbox{EuCd}_{2}\mbox{As}_{2}$, described in Section \ref{APPLA}, we find that the contribution $\delta \gv{\cal B}$, resulting with components  only in the direction of the separation  $\mathbf{\tilde b}$ of the nodes, lowers down the \textit{ab initio} calculations of the anomalous Hall conductivity  $\sigma _{yx} ^{\mbox{\scriptsize AHC}} \approx 30 \, \Omega ^{-1} \mbox{cm}^{-1}$  to the value  $\sigma _{yx} ^{\mbox{\scriptsize AHC}} \approx 15.9 \, \Omega ^{-1} \mbox{cm}^{-1}$
  which is much closer to the measured value of $0.5 \, \Omega ^{-1} \mbox{cm}^{-1}$. However there are other factors which could diminish further the conductivity such as  finite temperature effects and higher-order terms in the model Hamiltonian, which are not considered in our calculations. 
  
  The second application in Section \ref{APPLB} concerns $\rm{TaAs}$, the best known WSM.  Here we have the proliferation of 24 Weyl nodes, split in two families, 8 of them, denoted by W1, located on the $k_z=2\pi/c$ plane,  together with the remaining 16 W2 nodes placed away from this plane. The separation of each pair of nodes is along the $k_x$ direction such that only the ideal anomalous Hall conductivity $\sigma^{\rm AHC}_{zy}$ should be nonzero. Nevertheless, tilting and anisotropy  endow the correction $\delta\gv{\cal B}$ with components along the three directions yielding nonzero contributions to $\sigma^{\rm AHC}_{yx}$ and $\sigma^{\rm AHC}_{xz}$, clearly distinguished from the conventional anomalous Hall conductivity, and  which could in principle be measured. Finally, it is worth mentioning that our analysis can  be extended to other WSM materials of the TaAs family, such as TaP, NbAs and NbP, which are also highly anisotropic and tilted,  which nevertheless  present additional challenges  not fully  covered in this work.
  }

\newpage

\acknowledgments

A.G., A.M.-R, R.M.vD. and L.F.U. acknowledge support from the project CONACyT (M\'exico) \# CF-428214. A.M.-R. has been partially supported by DGAPA-UNAM Project No. IA102722.
Useful discussions with J. E. Barrios and E. Mu\~noz are greatly appreciated.

\appendix

\section{ The projection of the propagators} 

\label{APA}

To evaluate the trace appearing in Eq. (\ref{17}) for the vacuum polarization tensor we have to know the action of the projection operator $P _{\chi}$ to the right of the propagator $S(k)$. In the case of a massless standard  fermion this is a simple task, since the propagator $ i / k_{\mu} \gamma ^{\mu}$  can be readily rationalized as $i k_{\mu} \gamma ^{\mu} / k ^{2}$. However in the problem at hand, the presence of the $\gamma ^{5}$ is the denominator of the propagator requires a more subtle analysis, since the projectors do not have an inverse which could be directly applied to the denominator. To tackle this problem in a general frame, we introduce two numerical vectors $U_{\mu}$ and $V _{\mu}$, and define $(\slashed{U} - \gamma ^{5} \slashed{V}) ^{-1}$, with the usual slash notation $\slashed{U} = U _{\mu} \gamma ^{\mu}$. Assuming that $\slashed{U}$ has a inverse given by $\slashed{U} ^{-1} = \slashed{U} / U ^{2}$, with $U^{2} = U _{\mu} U ^{\mu}$, we consider the following sequence:
\begin{align}
    \frac{1}{\slashed{U} - \gamma ^{5} \slashed{V}} = \frac{1}{ 1 - \frac{ \slashed{U} }{U ^{2}} \gamma ^{5} \slashed{V}} \frac{ \slashed{U} }{U ^{2}} = \sum _{n} \left( \frac{ \slashed{U} }{U ^{2}} \gamma ^{5} \slashed{V} \right) ^{n} \frac{ \slashed{U} }{U ^{2}} = \frac{ \slashed{U} }{U ^{2}} \sum _{n} \left( \frac{ 1 }{U ^{2}} \gamma ^{5} \slashed{V} \slashed{U} \right) ^{n} . \label{General_Operator}
\end{align}
Now, we have to apply this expression to the projector $P _{\chi}$. To this end we use the identities $\gamma ^{\mu} P _{\chi} = P _{-\chi} \gamma ^{\mu} $ and $\gamma ^{5} \gamma ^{\mu} P _{\chi} = - \gamma ^{\mu} \gamma ^{5} P _{\chi} = -\chi \gamma ^{\mu} P _{\chi} $, such that
\begin{align}
    \gamma ^{5} \slashed{V} \slashed{U} P _{\chi} = \gamma ^{5} V _{\mu} \gamma ^{\mu} U _{\nu} \gamma ^{\nu} P _{\chi} = \gamma ^{5} V _{\mu} \gamma ^{\mu} P _{- \chi} U _{\nu} \gamma ^{\nu}  = \chi  V _{\mu} \gamma ^{\mu} U _{\nu} \gamma ^{\nu} = \chi P _{\chi} \slashed{V} \slashed{U} . \label{Idendity_Slash_Projector}
\end{align}
In this way, using the relation (\ref{Idendity_Slash_Projector}) for each motion of the projector $P _{\chi}$ to the left of a product $\slashed{V} \slashed{U}$ we obtain
\begin{align}
    \frac{1}{\slashed{U} - \gamma ^{5} \slashed{V}} P _{\chi}  = \frac{ \slashed{U} }{U ^{2}} \sum _{n} \left( \frac{ 1 }{U ^{2}} \gamma ^{5} \slashed{V} \slashed{U} \right) ^{n} P _{\chi} = \frac{ \slashed{U} }{U ^{2}}  P _{\chi}\sum _{n} \left( \frac{ 1 }{U ^{2}} \chi \slashed{V} \slashed{U} \right) ^{n} = P _{- \chi}\sum _{n} \frac{ \slashed{U} }{U ^{2}}   \left( \frac{ 1 }{U ^{2}} \chi \slashed{V} \slashed{U} \right) ^{n} ,
\end{align}
wherefrom we read the identity
\begin{align}
    \frac{1}{\slashed{U} - \gamma ^{5} \slashed{V}} P _{\chi}  =  P _{- \chi} \frac{1}{\slashed{U} - \chi \slashed{V}}  . \label{Identity_Propagator}
\end{align}
With the help of this result we can now evaluate the quantity $S(k) \gamma ^{\mu} P _{\chi}$, where $S(k) = i (\Gamma ^{\mu} k _{\mu} - M) ^{-1}$ is the fermion propagator with $\Gamma ^{\mu} = {\gamma^\mu}+c ^{\mu} _{\phantom{\mu} \nu} \gamma ^{\nu} + d ^{\mu} _{\phantom{\mu} \nu} \gamma ^{5} \gamma ^{\nu}$ and $M = a _{\mu} \gamma ^{\mu} + b _{\mu} \gamma ^{5} \gamma ^{\mu}$. In this way
\begin{align}
    S(k) \gamma ^{\nu} P _{\chi} = \frac{i}{\Gamma ^{\mu} k _{\mu} - M} P _{- \chi} \gamma ^{\nu} \equiv P _{\chi} S _{\chi} (k)   \gamma ^{\nu}  ,
\end{align}
where
\begin{align}
    S _{\chi} (k) = \frac{i}{ \left[  k _{\mu} (m _{\chi} ) ^{\mu} _{\phantom{\mu} \nu} - (C _{\chi}) _{\nu} \right] \gamma ^{\nu}}  \label{Chiral_Propagator}
\end{align}
is interpreted as the propagator for a fermion of chirality $\chi$ with
\begin{align}
    (m _{\chi} ) ^{\mu} _{\phantom{\mu} \nu} ={{\delta^\mu}_\nu}+c ^{\mu} _{\phantom{\mu} \nu} - \chi d ^{\mu} _{\phantom{\mu} \nu} , \qquad (C _{\chi}) _{\nu} = a _{\nu} - \chi b _{\nu} .  \label{MC}
\end{align}

\section{Calculation of the polarization tensor} 

\label{APB}

The goal of this Section is to evaluate the vacuum polarization of a fermion with chirality $\chi$, as defined by Eq. (\ref{21}). Substituting the chiral propagator (\ref{Chiral_Propagator}) into Eq. (\ref{21}) one gets 
\begin{align}
    i \Pi ^{\mu \nu } _{\chi} (p) = e ^{2} \, (m _{\chi}) ^{\mu} _{\phantom{\mu}\beta} (m _{\chi}) ^{\nu} _{\phantom{\mu}\alpha} \int \frac{d ^{4}k}{\left( 2 \pi \right) ^{4}} \, {\rm tr}   \left\lbrace \frac{i}{ \left[  (k _{\lambda} - p _{\lambda} ) (m _{\chi})^{\lambda}  _{\phantom{\lambda} \tau} - (C _{\chi}) _{\tau}   \right] \gamma ^{\tau} } \gamma ^{\beta}  \frac{i}{ \big[  k _{\sigma} \, m ^{\sigma} _{\phantom{\sigma} \xi } - ( C _{\chi} ) _{\xi} \big] \gamma ^{\xi} } \gamma ^{\alpha} \, P_\chi \right\rbrace . \label{Pol_Tensor_Aux}
\end{align}
Since our main concern in this paper is to calculate the CPT-odd contribution to the effective action of a general WSM, we now retain only the axial part of the {vacuum} polarization tensor, which arises from selecting the $\chi \gamma ^{5}/2$ part of the projector {in  the right of Eq. (\ref{Pol_Tensor_Aux})}. To simplify further such expression, we introduce the change of variables $k _{\nu}^{\prime } = k _{\mu}(m _{\chi}) ^{\mu} _{\phantom{\mu}\nu}$ such that $d ^{4} k = \frac{1}{\det m _{\chi} } d ^{4} k ^{\prime}$ and define $p _{\nu}^{\prime } = p _{\mu}(m _{\chi}) ^{\mu} _{\phantom{\mu}\nu}$. For simplicity in the notation we do not explicitly write the $\chi$-dependence of the primed variables $k _{\nu}^{\prime }$ and $p _{\nu}^{\prime }$. All in all, Eq. (\ref{Pol_Tensor_Aux}) becomes
\begin{align}
    i \Pi ^{\mu \nu } _{\chi} (p) = \frac{\chi}{2} e ^{2} \, (m _{\chi}) ^{\mu} _{\phantom{\mu}\beta} (m _{\chi}) ^{\nu} _{\phantom{\mu}\alpha} \frac{1}{\det m _{\chi} } \int \frac{d ^{4}k'}{\left( 2 \pi \right) ^{4}} \, {\rm tr}   \left[ \frac{i}{ \left(k ' - p ' - C _{\chi} \right) _{\tau}    \gamma ^{\tau} } \gamma ^{\beta}  \frac{i}{ \left(k ' - C _{\chi} \right) _{\xi}   \gamma ^{\xi} } \gamma ^{\alpha} \, \gamma ^{5} \right] . \label{Pol_Tensor_Aux_2}
\end{align}
Rationalizing the propagators and taking the trace using ${\rm tr} (\gamma ^{\tau }\gamma ^{\beta }\gamma ^{\xi }\gamma ^{\alpha }\gamma
^{5})=-4i\epsilon ^{\tau \beta \xi \alpha }$ with $\epsilon ^{0123}=+1$ we get 
\begin{align}
    i \Pi ^{\mu \nu } _{\chi} (p) &= - \frac{\chi}{2} e ^{2} \, (m _{\chi}) ^{\mu} _{\phantom{\mu}\beta} (m _{\chi}) ^{\nu} _{\phantom{\mu}\alpha} \frac{1}{\det m _{\chi} } \int \frac{d ^{4}k'}{\left( 2 \pi \right) ^{4}} \, {\rm tr}   \left[ \frac{ \left(k ' - p ' - C _{\chi} \right) _{\tau}     }{ \left(k ' - p ' - C _{\chi} \right) ^{2} }   \frac{ \left(k ' - C _{\chi} \right) _{\xi}  }{ \left(k ' - C _{\chi} \right) ^{2} } \, \gamma ^{\tau} \gamma ^{\beta}  \gamma ^{\xi} \gamma ^{\alpha}  \gamma ^{5} \right] \notag \\[5pt]  &= - 2\chi  e ^{2} \, (m _{\chi}) ^{\mu} _{\phantom{\mu}\beta} (m _{\chi}) ^{\nu} _{\phantom{\mu}\alpha} \frac{1}{\det m _{\chi} } \epsilon ^{\tau \beta \xi \alpha } \int \frac{d ^{4}k'}{\left( 2 \pi \right) ^{4}} \,  \frac{ p ' _{\tau} \left(k ' - C _{\chi} \right) _{\xi} }{ \left(k ' - p ' - C _{\chi} \right) ^{2}  \left(k ' - C _{\chi} \right) ^{2} }  ,  \label{Pol_Tensor_Aux_3}
\end{align}
which is the exact nonperturbative result. Since we are looking for the contribution to Eq. (\ref{BLAMBDA}) with only one power of the external momenta, we set $p'=0$ in the denominator. A further simplification arises due to the identity
\begin{align}
    m ^{\mu} _{\phantom{\mu} \alpha} m ^{\nu} _{\phantom{\nu}\beta} \epsilon ^{\alpha \beta \rho \sigma} p _{\rho} ^{\prime} =  p _{\kappa } \left( \det m _{\chi} \right) ( m ^{-1} _{\chi}) ^{\sigma} _{\phantom{\sigma}\lambda} \, \epsilon ^{\mu \nu \lambda \kappa }
\end{align}
yielding our final result of Eq. (\ref{23}), with the function $ I ^{\chi} _{\rho} (C)$ defined by Eq. (\ref{24}).

{
\section{The $\mu$-independent contribution}

\label{APint}
We now explicitly compute the integral (\ref{Nint})
\begin{eqnarray}
	N&=&\frac{1}{C_\chi^2(2\pi)^4} \int_{-\infty}^{\infty} dk'_0  \int_{-\infty}^{\infty} dk'_z  \int_0^{2\pi} d\phi \int_{0}^{\infty} dk'_{\perp} k'_{\perp} \frac{(k'_0-(C_\chi)_0) (C_\chi)_0+ (k'_z-(C_\chi)_z)(C_\chi)_z}{((k'_0-{(C_\chi)_0})^2+{k'_{\perp}}^2+(k'_z-(C_\chi)_z)^2)^2}\nonumber.
\end{eqnarray}
Here the integrals over $\phi$ and $k'_{\perp}$ are finite and can be immediately calculated. Therefore, after integrating, we obtain
\begin{equation}
		N_\chi=\frac{1}{2C_\chi^2(2\pi)^3}\left(I_0+I_z\right)
\end{equation}
where
\begin{eqnarray}
	I_0&=&(C_\chi)_0 \int_{-\infty}^{\infty} dk'_0 \int_{-\infty}^{\infty}dk'_z  \frac{(k'_0-(C_\chi)_0) }{(k'_0-{(C_\chi)_0})^2+(k'_z-{(C_\chi)}_z)^2},
\end{eqnarray}
and
\begin{eqnarray}
	I_z&=&(C_\chi)_z\int_{-\infty}^{\infty} dk'_0 \int_{-\infty}^{\infty}dk'_z  \frac{ (k'_z-{(C_\chi)}_z)}{(k'_0-{(C_\chi)_0})^2+(k'_z-{(C_\chi)}_z)^2}.
\end{eqnarray}
For $I_0$ the integral over $k'_z$ is finite, whereas the integral over $k_0$ shows a logarithmic divergence. Consequently, a cut-off is introduced in the $k'_0$ direction. To compute this integral, we initiate the process by integrating over $k'_z$ through the introduction of the substitution
\begin{equation}
(k'_z-(C_\chi)_z)^2 \rightarrow (k'_0-{(C_\chi)_0})^2\tan^2\theta \implies dk'_z=|k'_0-{(C_\chi)_0}|\sec^2\theta d\theta.
\end{equation}
Then, integrating over $\theta$, we arrive at
\begin{eqnarray}
	I_0&=&\pi(C_\chi)_0 \int_{-\Lambda_0}^{\Lambda_0} dk'_0 \mathrm{sgn}\left(  k'_0-(C_\chi)_0\right) \nonumber \\
	&=&\pi(C_\chi)_0\left( \int_{-\Lambda_0}^{(C_\chi)_0} dk'_0 \mathrm{sgn}\left(  k'_0-(C_\chi)_0\right)+\int_{(C_\chi)_0}^{\Lambda_0} dk'_0 \mathrm{sgn}\left(  k'_0-(C_\chi)_0\right) \right)\nonumber \\
	&=& \pi(C_\chi)_0\left[-\left( (C_\chi)_0+\Lambda_0\right)+\Lambda_0-(C_\chi)_0 \right]\nonumber \\
	&=&-2\pi(C_\chi)_0^2.
\end{eqnarray} 
This yields a finite result in which the divergences disappear.
For $I_z$, the situation is equivalent, with the difference that, in this case, the integral over $k'_0$ is finite while the integral over $k'_z$ displays a logarithmic divergence. Proceeding in analogous way to the previous case, we find that
\begin{equation}
		I_z=-2\pi(C_\chi)_z^2.
\end{equation}
So, finally, we obtain
\begin{equation}
		N=-\frac{1}{8\pi^2}.
\end{equation}
}

\section{Calculation of the poles and  residues of $g ^{\chi} _{\rho} (k _{0}, \gv{k})$ }  

\label{APC}

To evaluate the $\mu$-dependent contribution to the {vacuum} polarization tensor we have to calculate the integral (\ref{Integral_mu_dependent}), which requires first {to determine the poles in the variable $k_0$ of the function $g ^{\chi} _{\rho} (k _{0}, \gv{k})$ defined by Eq. (\ref{24}), together with the corresponding residues.}

\subsection{The poles}

Let us start by finding {the {positions $k ^{\chi \#} _{0 s}$ of the  double poles} of $g ^{\chi} _{\rho} (k _{0}, \gv{k})$ in the $k _{0}$-plane.}
From Eq. (\ref{24}) we see that they are located (in primed coordinates) at the two points $k ^{\prime} _{0 s} = (C _{\chi} )_{0} +s \vert {\gv k} ^{\prime} - {\gv C} _{\chi} \vert $. To find the corresponding $k ^{\chi \#} _{0 s}$ we recall the relations $k ^{\prime} _{0} = k _{\mu} (m _{\chi}) ^{\mu} _{ \phantom{\mu} 0}$ and $k ^{\prime} _{i} = k _{0} (m _{\chi}) ^{0} _{\phantom{0}i} + k _{j} (m _{\chi}) ^{j} _{\phantom{j}i}$, together with our general condition $(m _{\chi}) ^{0} _{\phantom{0}\nu} = \delta ^{0} _{\phantom{0}\nu}$  arising from   the linearized Hamiltonian (\ref{Hamiltonian_Tilting}). Under  this assumption, which avoids the mixing of $k _{0}$
and $k _{i}$  in $k'_{j}$, the poles are located at
\begin{align}
 k ^{\chi \#} _{0 s}= - k _{j} (m_\chi) ^{j}{}_{0} + (C_\chi) _{0} + s \,  \vert {\gv k} ^{\prime} - {\gv C} _{\chi} \vert ,   \label{POLES11}
\end{align}
{where $s= \pm 1$ denotes the band index.}
Using the equivalences in Eq. (\ref{SUMMARY}), one can further verify that the resulting  poles in the energy variable correspond exactly to the values of the energy obtained from the dispersion relation  (\ref{Energy}) calculated from the model Hamiltonian (\ref{Hamiltonian_Tilting}).
To see this, {let us start from Eq. (\ref{SUMMARY}) where }  we read that $(m _{\chi}) ^{i} _{\phantom{i}0} = {v} ^{i} _{\chi}$, $(m _{\chi}) ^{i} _{\phantom{i}j} = (A _{\chi}) _{ij}$, $(C _{\chi}) _{0} = \chi \,( { v} ^{i} _{\chi} \,  { {\tilde b}} ^{i} - {\tilde b} _{0})$ and $(C _{\chi}) _{j} = \chi \, {\tilde b} ^{i} (A _{\chi}) _{ij}$. {The substitution of these relations into the right hand side of Eq.(\ref{POLES11})} yields $k ^{\chi \#} _{0 s} = E _{s \chi}( {\gv k} )$.  

For later use in the main text we introduce double-primed momenta defined by the shifting ${\gv k} ^{\prime \prime} = {\gv k} ^{\prime} - {\gv C} _{\chi}$, such that the poles (\ref{POLES11}) can be written in the simple form
\begin{align}
 k ^{\chi \#} _{0 s}  = \gv{\mathcal{V}}_\chi \cdot {\gv k} ^{\prime \prime}  + s \, \vert {\gv k} ^{\prime \prime}  \vert + E_{\chi 0} , \label{POLES_D_primed}
\end{align}
where we define ${\cal V} _{\chi} ^{j} =(m _{\chi} ^{-1}) ^{j} _{\phantom{j}i} \, (m _{\chi}) ^{i} _{\phantom{i}0}$ and $E _{\chi 0} = {\gv{\cal V}} _{\chi} \cdot {\gv{C}} _{\chi} + (C _{\chi}) _{0}$. To derive this result we  use the sequence of relations: $k _{i} (m _{\chi}) ^{i} _{\phantom{i}0} = k' _{j} ( m _{\chi} ^{-1}) ^{j} _{\phantom{j}i} \, (m _{\chi}) ^{i} _{\phantom{i}0} = \left[  k''_{j} + (C _{\chi}) _{j} \right] (m _{\chi} ^{-1}) ^{j} _{\phantom{j}i} \, (m _{\chi} ) ^{i} _{\phantom{i}0} $. In the double-primed coordinates the gap closing condition (defining the nodes) is ${\gv k} ^{\prime \prime} = {\gv 0}$, such that $E _{\chi 0}$ corresponds to the position of the nodes in energy. As we explicitly verify in the Appendix \ref{APE}, {the  expression defined  for $E_{\chi 0}$ in Eq. (\ref{POLES_D_primed}) 
} reproduces the expected result  $E_{\chi 0}=- \chi{\tilde b} _{0} $.

\subsection{The residues}

For second order poles, the residue of $g ^{\chi} _{\rho} (k _{0}, \gv{k})$ in {the variable} $k _{0}$ is
\begin{equation}
    \mathrm{Res} \, [ g ^{\chi} _{\rho} (k _{0}, \gv{k}) ] = \frac{d}{dk _{0} } \left[  \left(k _{0} - k ^{\chi \#} _{0 s} \right) ^{2}    g ^{\chi} _{\rho} (k _{0} , \gv{k}) \right] \bigg| _{k _{0} = k ^{\chi \#} _{0 s}},
\end{equation}
where $k ^{\chi \#} _{0 s}$ {denote} the position of the poles, given by Eq. (\ref{POLES11}). To proceed, we first write $g ^{\chi} _{\rho} (k _{0}, \gv{k})$ as an explicit function of $k _{0}$ and $k _{i}$, i.e.
\begin{align}
    g ^{\chi} _{\rho} (k _{0}, \gv{k}) = \frac{ \left[ k _{0} + (m _{\chi}) ^{j} _{\phantom{j}0} \, k _{j} \right] \delta ^{0} _{\phantom{0} \rho } + ( m _{\chi} ) ^{j} _{\phantom{j}i} \, k _{j} \, \delta ^{i} _{\phantom{i} \rho} -(C _{\chi}) _{\rho}}{(k _{0} -  k _{0s} ^{\chi \#} ) ^{2} ( k _{0} - k _{0-s} ^{\chi \#} ) ^{2}}
\end{align}
Calculating the residue yields
\begin{align}
    \mathrm{Res} \, [ g ^{\chi} _{\rho} (k _{0}, \gv{k}) ]  &= \frac{d}{dk _{0} } \left\lbrace \frac{ \left[ k _{0} + (m _{\chi}) ^{j} _{\phantom{j}0} \, k _{j} \right] \delta ^{0} _{\phantom{0} \rho } + ( m _{\chi} ) ^{j} _{\phantom{j}i} \, k _{j} \, \delta ^{i} _{\phantom{i} \rho} - (C _{\chi}) _{\rho}}{ ( k _{0} - k _{0-s} ^{\chi \#} ) ^{2}} \right\rbrace \Bigg| _{k _{0} = k ^{\chi \#} _{0 s}}  \notag \\ &= \delta ^{0} _{\phantom{0} \rho } \left[ \frac{1}{ ( k _{0s} ^{\chi \#} - k _{0-s} ^{\chi \#} ) ^{2}} - 2  \frac{ k _{0s} ^{\chi \#} + (m _{\chi}) ^{j} _{\phantom{j}0} \, k _{j}  - (C _{\chi}) _{0} }{ ( k _{0s} ^{\chi \#} - k _{0-s} ^{\chi \#} ) ^{3}}  \right] - 2  \delta ^{i} _{\phantom{i} \rho } \frac{ (m _{\chi}) ^{j} _{\phantom{j}i} \, k _{j}  - (C _{\chi}) _{i} }{ ( k _{0s} ^{\chi \#} - k _{0-s} ^{\chi \#} ) ^{3} } .
\end{align}
However, from Eq. (\ref{POLES11}) one finds
\begin{align}
    k _{0s} ^{\chi \#} - k _{0-s} ^{\chi \#} =  2s \vert \gv{k}' - \gv{C} _{\chi} \vert , \qquad k _{0s} ^{\chi \#} + (m _{\chi}) ^{j} _{\phantom{j}0} \, k _{j} - (C _{\chi}) _{0} = s \vert \gv{k}' - \gv{C} _{\chi} \vert ,
\end{align}
and thus
\begin{align}
    \mathrm{Res} \, [ g ^{\chi} _{\rho} (k _{0}, \gv{k}) ]  &= \delta ^{0} _{\phantom{0} \rho } \left[ \frac{1}{ 4 \vert \gv{k}' - \gv{C} _{\chi} \vert ^{2}} - 2 \, \frac{ s \vert \gv{k}' - \gv{C} _{\chi} \vert }{ 8 s \vert \gv{k}' - \gv{C} _{\chi} \vert ^{3}}  \right] - 2  \delta ^{i} _{\phantom{i} \rho } \frac{ k' _{i}  - (C _{\chi}) _{i} }{  8 s \vert \gv{k}' - \gv{C} _{\chi} \vert ^{3} } =   -  s \delta ^{i} _{\phantom{i} \rho } \frac{ k' _{i}  - (C _{\chi}) _{i} }{ 4 \vert \gv{k}' - \gv{C} _{\chi} \vert ^{3} }   .
\end{align}
We observe that the residues have the monopole-like structure resembling the Berry curvature (\ref{BerryCurvature}).

\section{Calculation of  $\gv{I}^{\chi s }$} \label{APD}

In the main text we find twice the generic integral
\begin{align}
\gv{I} ^{\chi s} = \int \frac{d ^{3} \gv{q} }{(2 \pi )^{3}} \frac{\gv{q}}{|\gv{q}|^{3}} \; H ( \mu - Q ^{s} ), \label{ICHIS}
\end{align}
with $Q ^{s} =  \gv{W} \cdot \gv{q} + s \vert \gv{q} \vert + E _{0}$,  where $s=\pm 1$ denotes the band index. The dependence of $\gv{I}^{\chi s}$ on the chirality $\chi$ is implicit in the vector $\gv{W}$, related to the tilting and the anisotropy, together with  $E _{0}$  which determines the location in energy of each node. In the following we do not indicate the chirality index for simplicity in the notation. This integral naturally emerges when computing the $\mu$-dependent contribution to the vacuum polarization tensor in Section \ref{IVB} as well as in Section \ref{Kinetic_Theory_App} when we evaluate the anomalous Hall current in a kinetic theory approach.

For the subsequent analyses we choose a spherical coordinate system with $\gv{W}$ pointing along the $z$-axis, such that $Q ^{s} = s \vert \gv{q} \vert \left( 1 + s \vert \gv{W} \vert  \cos \theta \right) + E _{0}$. The Heaviside function $H$ appearing in the integral (\ref{ICHIS})  imposes the restriction
\begin{align}
\mu - E _{0} > s \vert \gv{q} \vert \left( 1 + s \vert \gv{W} \vert  \cos \theta \right) , \label{Restriction}
\end{align}
which {requires some care for its implementation} since it depends on the magnitude of $\vert \gv{W} \vert $. In this paper we consider the case $\vert \gv{W} \vert < 1$, which corresponds to a type-I WSM. With this choice, it is clear that $1 + s \vert \gv{W} \vert  \cos \theta > 0$ for all values of  $\theta$, and hence the band index controls the sign of the right hand side  of Eq. (\ref{Restriction}). On the other hand, the case  $\vert \gv{W} \vert > 1$, which corresponds to a type-II WSM, presents additional difficulties since the right hand side  changes sign at $\theta = \arccos (-s/ \vert \gv{W} \vert )$  and make some integrals diverge, thus requiring to introduce additional cutoffs in the model. This case is beyond the scope of the present work. 
The simplest way of calculating the integral (\ref{ICHIS}) is to take advantage of the identity
\begin{align}
\frac{ {\gv{q}} }{ \vert {\gv{q}} \vert ^{3} } = - \nabla _{ q } \frac{1}{ \vert {\gv{q}} \vert },
\end{align}
and subsequently integrate by parts to obtain
\begin{align}
\gv{I} ^{\chi s} = - \int \frac{d ^{3} \gv{q} }{(2\pi )^{3}}  \left\lbrace \nabla
_{q} \left[ \frac{1}{ \vert {\gv{q}} \vert } \, H ( \mu -Q ^{s}) \right] -\frac{1}{ \vert {\gv{q}} \vert }\nabla _{q} \, H(\mu -Q^{s}) \right\rbrace .  \label{Integral_I_appendix}
\end{align}
{Using the Gauss theorem, the first term (to be called $\gv{I} ^{\chi s} _{S}$) yields the surface integral }

\begin{align}
\gv{I} ^{\chi s} _{S} = - \lim _{ \vert {\gv{q}} \vert \to \infty } \oint  \frac{ d \gv{S} }{ (2 \pi ) ^{3} }  \frac{1}{ \vert {\gv{q}} \vert } \, H ( \mu -Q ^{s}).  \label{Surface_Int}
\end{align}
To evaluate this integral we choose a spherical-shaped surface centered at the origin such that $d \gv{S} = \hat{{\gv{q}}} \, d \Omega $, where $d \Omega = \sin \theta d \theta d \phi$ is the differential of solid angle. Besides, we observe that we require the limit of the Heaviside function when $\vert {\gv{q}} \vert \to \infty$. For type-I WSMs ($\vert \gv{W} \vert < 1 $) we find
\begin{align}
\lim _{ \vert {\gv{q}} \vert \to \infty } H ( \mu -Q ^{s}) = H (-s) , 
\end{align}
since $1 + s \vert \gv{W} \vert  \cos \theta > 0$ for all values of  $\theta$. This makes the integrand of Eq. (\ref{Surface_Int}) independent of the angular variables, thus implying 
\begin{align}
\gv{I} ^{\chi s} _{S} = - H (-s) \oint  \frac{ d \Omega }{ (2 \pi ) ^{3} }  \frac{ \hat{{\gv{q}}}}{ \vert {\gv{q}} \vert }  = 0 .
\end{align}
Thus we are left with the second term of Eq. (\ref{Integral_I_appendix}), which we rewrite as follows:
\begin{align}
\gv{I} ^{\chi s} =   \int \frac{d ^{3} \gv{q} }{(2\pi )^{3}}  \frac{1}{ \vert {\gv{q}} \vert }\nabla _{q} \, H(\mu -Q^{s}) = - \int \frac{d ^{3} \gv{q} }{(2\pi )^{3}}  \frac{ \nabla _{q} Q^{s} }{ \vert {\gv{q}} \vert }  \, \delta (\mu -Q^{s})  . \label{Integral_I_appendix_2}  
\end{align}
Now, using the fact that $\nabla _{q} Q ^{s} =  \gv{W} + s \hat{{\gv{q}}} $ and introducing $\Lambda \equiv \mu -E _{0}$, we obtain
\begin{align}
\gv{I} ^{\chi s} =  - \int \frac{d ^{3} \gv{q} }{(2\pi )^{3}}  \frac{ \gv{W} + s \hat{{\gv{q}}} }{ \vert {\gv{q}} \vert }  \, \delta \left[ \Lambda - s \vert \gv{q} \vert (1 + s \gv{W} \cdot \hat{{\gv{q}}} ) \right]  .  
\end{align}
Since the only vector at our disposal is $\gv{W}$ we have $\gv{I}^{\chi s} = M ^{\chi s} \gv{W}$, where
\begin{align}
M ^{\chi s} = - \frac{1}{W ^{2} } \int \frac{d ^{3} \gv{q} }{ (2 \pi )^{3} }  \frac{  \gv{W} \cdot ( \gv{W} + s \hat{{\gv{q}}} ) }{ \vert {\gv{q}} \vert }  \, \delta \left[ \Lambda - s \vert \gv{q} \vert (1 + s \gv{W} \cdot \hat{{\gv{q}}} ) \right] . \label{DEFN}
\end{align}
To evaluate this integral we employ a spherical coordinate system with  $\gv{W}$ pointing along the $z$-axis. The radial integral can be performed by decomposing the Dirac delta as
\begin{align}
\delta \left[ \Lambda - s \vert \gv{q} \vert (1 + s \gv{W} \cdot \hat{{\gv{q}}} ) \right] = \frac{ \delta ( \vert \gv{q} \vert - q ^{\ast}  ) }{ \vert 1 + s \gv{W} \cdot \hat{{\gv{q}}} \vert } H (q ^{\ast} ) ,  \label{Decomp_Delta}
\end{align}
where
\begin{align}
q ^{\ast} = \frac{s \Lambda}{  1 + s \gv{W} \cdot \hat{{\gv{q}}}} .  \label{Root}
\end{align}
Note that the Heaviside function in Eq. (\ref{Decomp_Delta}) guarantees that the root (\ref{Root}) should be positive. The case at hand is simple since $\vert  \gv{W} \vert < 1 $ implies $1 + s \gv{W} \cdot \hat{{\gv{q}}} > 0$ and therefore the Heaviside function in (\ref{Decomp_Delta}) restricts the product $s \Lambda $ to be positive. Therefore, performing the radial integration in (\ref{DEFN}) we obtain
\begin{align}
M ^{\chi s} = - \frac{s \Lambda}{ \vert  \gv{W} \vert }  H (s \Lambda) \, \int \frac{d \Omega }{ (2 \pi )^{3} }  \frac{  \vert  \gv{W} \vert + s \cos \theta }{ ( 1 + s \vert  \gv{W} \vert \cos \theta ) ^{2} }  . 
\end{align}
This integral can be easily computed with a simple change of variables. The final result is 
\begin{align}
M ^{\chi s} =  \frac{s \Lambda}{2 \pi ^{2} \vert  \gv{W} \vert ^{3} }  H (s \Lambda)  \; \left( \vert  \gv{W} \vert -  {\rm arctanh} \vert  \gv{W} \vert  \right) ,
\end{align}
such that $\gv{I}^{\chi s} = M ^{\chi s} \gv{W}$ .

\section{The vector ${\cal B}_\lambda$} \label{APE}

Following Eqs. (\ref{DEFL}) and (\ref{BLAMBDA}) we realize that the effective action is determined by the vector ${\cal B} _{\lambda}$, which has a contribution given by Eq. (\ref{29a}), to be called universal for the reasons indicated after obtaining ${\cal B} ^{(1)}_{\lambda}$ in this section,   together with a correction term due to finite density given by Eq. (\ref{BLAMBDA2}). Our equations (\ref{29a}) and (\ref{BLAMBDA2}) are expressed in terms of the parameters of the SME. The goal of this section is to rewrite the vector ${\cal B} _{\lambda}$ in terms of the parameters describing the condensed matter Hamiltonian (\ref{Hamiltonian_Tilting}).

Our first task is to compute the inverse of the matrix $(m _{\chi}) ^{\mu} _{\phantom{\mu}\nu}$. To this end we use the representation (\ref{Convenciones_V513}) together with the equivalences in Eq. (\ref{SUMMARY}). In this case, the matrix $[(m _{\chi}) ^{\mu} _{\phantom{\mu}\nu}]$ can be written as
\begin{align}
    [(m_\chi)^{\mu}{}_\nu ] \equiv \begin{pmatrix} 1 &\,\,  0 \\ \gv{v}_\chi & \,\, \mathbb{A}_\chi \\    \end{pmatrix}, \qquad \mathbb{A}_\chi=[(A_\chi)_{ij}] , \label{E1}
\end{align}
such that the inverse is given by
\begin{align}
    [(m^{-1}_\chi)^{\mu}{}_\nu] \equiv \begin{pmatrix} 1 & \,\, 0 \\    -{\mathbb A}_\chi^{-1}    \gv{v} _{\chi} & \,\,\,\, {\mathbb A}_\chi^{-1}    \\    \end{pmatrix}, \qquad     (A_\chi)_{ij} \, (A^{-1}_\chi)_{jk}=\delta_{ik}.    \label{E2}
\end{align}
Therefore, the required components (i.e. space-time and space-space) of this matrix become
\begin{align}
    {(m _{\chi} ^{-1})} ^{i} _{\phantom{i} 0} = -[{\mathbb A} _{\chi} ^{-1} \gv{v} _{\chi}] _{i} = - (A ^{-1} _{\chi} ) _{ij}\, v _{\chi} ^{j} , \qquad  (m _{\chi} ^{-1}) ^{i} _{\phantom{i}j} = (A ^{-1} _{\chi}) _{ij} .   \label{E3}
\end{align}
Besides the above inverses we also need ${\cal V} ^{i}$, which is given by Eq. (\ref{ADDDEF}). Note that ${\cal V} ^{i}$ yields $N _{\chi}$ in Eq. (\ref{NCHI}). Substituting equations (\ref{E1}) and (\ref{E3}) into Eq. (\ref{ADDDEF}) we obtain
\begin{align}
    {\cal V _{\chi} } ^{j} = ( m _{\chi} ^{-1} ) ^{j} _{ \phantom{j} i} \,  (m _{\chi}) ^{i} _{\phantom{i} 0} = (A _{\chi} ^{-1})_{ji} v _{\chi} ^{i} . \label{V_app}
\end{align}
Now we have the missing ingredients to establish the correspondence between the SME coefficients with the parameters appearing in the condensed matter Hamiltonian (\ref{Hamiltonian_Tilting}).

On the one hand, using $(C _{\chi}) _{0} = \chi \,( { v}^{i} _{\chi} \,  { {\tilde b} ^{i} }  -  {\tilde b}_{0}) $ and $(C _{\chi}) _{j} = \chi \, {\tilde b} ^{i} (A _{\chi}) _{ij}$ defined by Eq. (\ref{SUMMARY}), we compute the vector ${\cal B}^{(1)} _{\lambda}$ given by Eq. (\ref{29a}) which determines the universal contribution. Taking our choice $N = - 1/(8 \pi ^{2})$ the time component becomes
\begin{align}
    \mathcal{B} _{0} ^{(1)} &= - \frac{1}{2} \sum _{\chi = \pm 1} \chi \left[  (C _{\chi}) _{0} \, (m _{\chi} ^{-1})^{0} _{ \phantom{\rho}0} + (C _{\chi}) _{j} \, (m _{\chi} ^{-1})^{j} _{ \phantom{j}0} \right] \notag \\ &= - \frac{1}{2} \sum _{\chi = \pm 1} \chi \left[ \chi \,( { v}^{i} _{\chi} \,  { {\tilde b} ^{i} }  -  {\tilde b}_{0})  - \chi \, {\tilde b} ^{i} (A _{\chi}) _{ij} \, (A ^{-1} _{\chi} ) _{jk}\, v _{\chi} ^{k} \right] \notag \\ &= - \frac{1}{2} \sum _{\chi = \pm 1}   \left[  ( { v}^{i} _{\chi} \,  { {\tilde b} ^{i} }  -  {\tilde b}_{0})  -  {\tilde b} ^{i} \, { v}^{i} _{\chi} \right] = {\tilde b}_{0} , 
\end{align}
while the space-component simplifies to
\begin{align}
    \mathcal{B} _{i} ^{(1)} &= - \frac{1}{2} \sum _{\chi = \pm 1} \chi \left[  (C _{\chi}) _{0} \, (m _{\chi} ^{-1})^{0} _{ \phantom{0}i} + (C _{\chi}) _{j} \, (m _{\chi} ^{-1})^{j} _{ \phantom{j}i} \right] \notag \\ &= - \frac{1}{2} \sum _{\chi = \pm 1} \chi \left[ 0 + \chi \, {\tilde b} ^{k} (A _{\chi}) _{kj} \, (A ^{-1} _{\chi}) _{ji} \right] \notag \\ &= - \frac{1}{2} \sum _{\chi = \pm 1} {\tilde b} ^{i}   = {\tilde b}_{i} . 
\end{align}
Note that ${\cal B}^{(1)} _{\lambda}$ depends only upon the position of the Weyl nodes in momentum and energy, and in this sense we call this contribution universal, since it becomes independent of the anisotropy, tilting and chemical potential.

On the other hand, using Eq. (\ref{V_app}) we obtain
\begin{align}
    E _{\chi 0} = {\gv{\cal V}} _{\chi} \cdot {\gv{C}} _{\chi} + (C _{\chi}) _{0} = (A _{\chi} ^{-1})_{ji} v _{\chi} ^{i} \left[ \chi\, {\tilde b}^{k} (A _{\chi})_{kj} \right] + \chi \,( { v}^{i} _{\chi} \,  { {\tilde b} ^{i} }  -  {\tilde b}_{0}) = -\chi \, {\tilde b}_{0} , 
\end{align}
such that $\Lambda _{\chi} = \mu + \chi {\tilde b} _{0} $. This is a measure of the position of the nodes with respect to the chemical potential. Now, we evaluate the components of the vector ${\cal B}^{(2)} _{\lambda}$, given by Eq. (\ref{BLAMBDA2}), which determines the $\mu$-dependent contribution. The time component becomes
\begin{align}
    {\cal B}^{(2)} _{0} &= - \sum _{\chi = \pm 1} \chi \Lambda _{\chi} N _{\chi} (m _{\chi} ^{-1}) ^{j} _{\phantom{j} 0} ({\cal V} _{\chi}) _{j} \notag \\ &= - \sum _{\chi = \pm 1} \chi \Lambda _{\chi} N _{\chi} \left[ -(A _{\chi} ^{-1} ) _{ji} v _{\chi} ^{i} \right] \left[ (-A ^{-1} _{\chi}) _{jk}\, v _{\chi} ^{k} \right] \notag \\ &= - \sum _{\chi = \pm 1} \chi \Lambda _{\chi} N _{\chi} \, {\gv{v}} _{\chi} \left( \mathbb{A} _{\chi} ^{-1} \right) ^{T} \mathbb{A} _{\chi} ^{-1} {\gv{v}} _{\chi}   \label{0B2} ,
\end{align}
while the space-component simplifies to
\begin{align}
    {\cal B}^{(2)} _{i} &= - \sum _{\chi = \pm 1} \chi \Lambda _{\chi} N _{\chi} (m _{\chi} ^{-1}) ^{j} _{\phantom{j} i} ({\cal V} _{\chi}) _{j} \notag \\ &=  - \sum _{\chi = \pm 1} \chi \Lambda _{\chi} N _{\chi} \, \left(  \mathbb{A}_{\chi} ^{-1}\right) _{ji}  \left[  - \left( \mathbb{A} _{\chi} ^{-1}\right) _{jk} v _{\chi} ^{k} ) \right] \notag \\ &= + \sum _{\chi = \pm 1} \chi \Lambda _{\chi} N _{\chi} \, \left[ \left(  \mathbb{A} _{\chi} ^{-1} \right) ^{T} \mathbb{A} _{\chi} ^{-1} {\gv{v}} _{\chi} \right] _{i} .
\end{align}
In these expressions 
\begin{align}
    N _{\chi} =  \frac{ 1 }{2 \vert {\gv{\cal V}} _{\chi} \vert ^{3} } \left( \vert {\gv{\cal V}} _{\chi} \vert - {\rm arctanh}\left(\vert {\gv{\cal V}} _{\chi} \vert \right)\right) ,
\end{align}
where ${\gv{\cal V}} _{\chi} = {\mathbb A} ^{-1} _{\chi} \gv{v}_\chi$. All in all, we have obtained the full characterization of the effective action $S _{\rm eff} ^{(2)} (A)$ given by Eq. (\ref{DEFL}) in terms of the polarization tensor $\Pi^{\mu\nu}$ defined by Eq. (\ref{BLAMBDA}), with the vector ${\cal B} _{\lambda}$ written in terms of the parameters of the  Hamiltonian of the WSM (\ref{Hamiltonian_Tilting}).

Summarizing, the  results are 
\barr
&&{\cal B}_0= {\tilde b}_0 -\sum_{\chi=\pm 1} \chi \Lambda_\chi \, N_\chi \, \gv{v}_\chi ({\mathbb A}_\chi^{-1})^T  ({\mathbb A}_\chi^{-1}) \gv{v}_\chi,    
\nonumber \\
&&{\cal B}_k= {\tilde b}_k+ \sum_{\chi=\pm 1} \chi \Lambda_\chi \, N_\chi \, 
\left[({\mathbb A}_\chi^{-1})^T ({\mathbb A}_\chi^{-1}) 
\gv{v}_\chi \right]_k, 
\nonumber \\
&& \Lambda_\chi= \mu + \chi{\tilde b}_0,  \qquad 
{\gv{\cal V}}_\chi= {\mathbb A}^{-1}_\chi \gv{v}_\chi, \qquad  N_\chi=\frac{1}{2 |{\gv{\cal V}}_\chi|^3} \left( |{\gv{\cal V}}_\chi|-{\rm arctanh}(|{\gv{\cal V}}_\chi|) \right).
\label{GENREL}
\earr

\bibliography{Referencias}
\end{document}